\documentclass[pdflatex,sn-mathphys-num]{sn-jnl}

\usepackage[T1]{fontenc}
\usepackage{graphicx}%
\usepackage{multirow}%
\usepackage{amsmath,amssymb,amsfonts}%
\usepackage{amsthm}%
\usepackage{mathrsfs}%
\usepackage[title]{appendix}%
\usepackage{xcolor}%
\usepackage{textcomp}%
\usepackage{manyfoot}%
\usepackage{booktabs}%
\usepackage{algorithm}%
\usepackage{algorithmicx}%
\usepackage{algpseudocode}%
\usepackage{listings}%
\usepackage[version=4]{mhchem}
\usepackage{siunitx}

\DeclareSIUnit{\curie}{Ci}

\theoremstyle{thmstyleone}%
\theoremstyle{thmstyletwo}%

\theoremstyle{thmstylethree}%

\begin{document}

\title[Low-temperature Performance of GAGG:Ce]{Low-temperature Performance of $\mathrm{Gd_3(Ga, Al)_5O_{12}}$:Ce Scintillators}


\author*{\fnm{Merlin} \sur{Kole}}\email{Merlin.Kole@unh.edu}

\author{\fnm{Kasun} \sur{Wimalasena}}\email{IllawathanneGedaraKasun.Wimalasena@unh.edu}

\author{\fnm{Richard} \sur{Gorby}}

\author{\fnm{Torsten} \sur{Diesel}}

\author{\fnm{Zachary} \sur{Greenberg}}

\author{\fnm{Fabian} \sur{Kislat}}

\affil{\orgdiv{Department of Physics \& Astronomy and Space Science Center}, \orgname{University of New Hampshire}, \orgaddress{\street{8 College Rd}, \city{Durham}, \state{NH} \postcode{03824}, \country{USA}}}

\abstract{The last years have seen the first cryogenic detectors to be proposed for usage on balloon-borne missions. In such missions, the instrument will be exposed to the high radiation environment of the upper atmosphere. This radiation can induce a significant background to the measurements, something which can be mitigated through the use of an anti-coincidence shield. For hard X-ray and gamma-ray detectors such a shield typically consists photomultiplier tubes or, more recently, silicon photomultipliers coupled to scintillators placed around the detector. When using cryogenic detectors, the shield can be placed around the entire cryostat which will make it large, heavy and expensive. For the ASCENT (A SuperConducting ENergetic x-ray Telescope) mission, which uses Transition Edge Sensor microcalorimeter detectors, it was therefore considered to instead place the shield inside. This comes with the challenge of operating it at cryogenic temperatures. For this purpose, we tested the performance of 2 different types of GAGG:Ce scintillators down to 15 mK for the first time. Although significant variations of both the decay time and the light yield were found when varying the temperature, at 4 K its performance was found to be similar to that at room temperature. Furthermore, unexpected behavior around 2 K was found for both types of GAGG:Ce, leading to more in depth studies around these temperatures. Overall, the studies show that the combination of materials will allow to produce a functional anti-coincidence shield at several Kelvin.}


\maketitle

\section{Introduction}\label{sec:intro}

Transition Edge Sensor (TES) microcalorimeter X-ray and gamma-ray detectors achieve an energy resolution better than \qty{0.1}{\percent}~\cite{WOLLMAN2000145,2022arXiv221006914G}, an order of magnitude better than state-of-the-art semiconductor detectors.
This makes them an attractive option for future astrophysical observatories.
For example, \textit{NewAthena} will use a microcalorimeter array for its X-ray Integrated Field Unit sensitive in the \qtyrange[range-phrase=--]{0.2}{12}{\keV} energy range~\cite{Barret2023, ESA_NewAthena}.
In the \qtyrange[range-phrase=--]{20}{80}{keV} hard X-ray band, the proposed balloon-borne telescope \textit{ASCENT} (A SuperConducting ENergetic x-ray Telescope) will employ a 512-pixel gamma-ray TES array with an energy resolution of \qty{67}{\eV} as the focal plane instrument of a hard X-ray telescope~\cite{kislat_etal_2023}.
At even higher energies, the proposed balloon-borne \textit{511-CAM} (\qty{511}{keV} gamma-ray camera using microcalorimeters) mission targets an energy resolution of \qty{390}{\eV} at \qty{511}{\keV}~\cite{shirazi_etal_2023}.
The unprecedented energy resolution of TES detectors comes with the complexity that they need to be operated at sub-Kelvin temperatures.

Gamma-ray detectors in space, or on a stratospheric balloon, will not only be exposed to photons from the source but also to the particles which make up their respective radiation environment.
Without shielding, these will induce a significant background to the measurements~\cite{iyer_etal_2023, Metzler:2024dgk}.
Typically, a combination of passive and active shielding will be employed. Active shielding consists of a secondary particle detector surrounding the focal plane instrument. Anti-coincidence detector (ACD) is form of active shielding capable of rejecting signals produced by background particles even if they only deposit part of their energy in the ACD. Therefore, an ACD can achieve the same background rejection as passive shielding with a significantly lower mass.

%

For soft X-ray missions the main source of background is expected to come from charged particles, especially for in-orbit missions.
Sufficient background rejection can therefore be achieved by placing a thin silicon-based detector behind the focal plane array, as developed for the cryogenic anti-coincidence for \textit{NewAthena}~\cite{d-andrea_etal_2024}.
For the \textit{Chandra} mission the same goal was achieved using a plastic scintillator with PMT readout~\cite{Murray2000}.
Hard X-ray and gamma-rays will be capable of penetrating silicon detectors, which means that instruments aiming to measure at these energy bands require ACDs with a higher stopping power.
This can be achieved by using high density and high-Z scintillators connected to photo-multiplier tubes (PMTs), such as that employed successfully on the SPectrometer aboard INTEGRAL (SPI) mission onboard INTErnational Gamma-Ray Astrophysics Laboratory (\textit{INTEGRAL})~\cite{SPI-ACS}, the Polarised Gamma-ray Observer (\textit{PoGOLite}) mission \cite{MariniBettolo2008}, the XL-Calibur mission \cite{iyer_etal_2023}, or the Nuclear Spectroscopic Telescope Array (\textit{NuSTAR}) \cite{Grefenstette2022}.
In recent years PMTs, which are bulky and fragile, are often being replaced in ACD designs by silicon photomultipliers (SiPMs), an example being that of the upcoming High Energy cosmic Radiation Detection (\textit{HERD}) detector~\cite{HERD_ACS}.

This shielding requirement poses a new challenge for cryogenic gamma-ray detectors.
Operation of the ACD at ambient temperature would require fully enclosing the entire cryostat, which is often not feasible from both a mass and a cost perspective.
In the original design of \textit{ASCENT}~\cite{kislat_etal_2023} the ACD only partially surrounds the focal plane instrument, significantly reducing its background rejection efficiency.
Here, we evaluate the feasibility of an ACD consisting of high-density, high-$Z$ inorganic scintillator at the \qty{4}{K} stage of the cryostat, making it possible to fully enclose the focal plane array.
This requires that both the scintillator and the SiPM or PMT have to operate at~\qty{4}{\K}. New SiPMs by Fondazione Bruno Kessler (FBK) of the type 'NUV-HD-Cryo' have been tested to perform down to \qty{10}{K} \cite{FBK, 2023arXiv231110497G} while additional studies by us, to appear in a future publication, also indicate they operate well down to \qty{4}{K}. 
In the present paper we focus on the choice of scintillator material and its cryogenic performance.
Readout of this scintillator with a the NUV-HD-Cryo SiPM operated at \qty{4}{\K} will be the subject of a future paper.


The ACD of previous missions, such as \textit{INTEGRAL} SPI~\cite{SPI-ACS} and NuSTAR employs \ce{Bi4Ge3O12} (BGO) scintillators and CsI, respectively.
The light yield of BGO is relatively low at around 10 optical photons per keV.
However, its high density of \qty{7.12}{\g/\cubic\cm} and high effective atomic number of 74 provides it with a large stopping power for both photons and charged particles.
Furthermore, its non-hygroscopic nature makes it relatively easy to produce large fully enclosing shielding structures without the need for passive encasing materials.
One downside of BGO is its relatively long decay time, typically of the order of $\mu s$. Furthermore, it exhibits a significant temperature-dependence of its performance.
While the light yield increases by a factor of about 5 when decreasing from room temperature to \qty{10}{\K}, the decay time of the scintillation pulse increases by 2 orders of magnitude to over \qty{100}{\micro\s}~\cite{BGO_temp,PhysRevB.84.214306}.
This would result in significant pulse pile-up when operated on a stratospheric balloon or in low-Earth orbit.
Thallium doped CsI is faster than BGO with a decay time of of around \qty{1}{\micro\s}.
However, its stopping power is lower than BGO due to its lower effective atomic number (54) and its density of only \qty{4.51}{\g/\cubic\cm}.
Similar to BGO, the performance of CsI at colder temperatures is significantly worse than at room temperature.
The decay time increases by a factor 3 at \qty{77}{\K} while its light output decreases by a factor 15~\cite{SWIDERSKI201932}.

Cerium-doped $\mathrm{Gd_3(Ga, Al)_5O_{12}}$ (GAGG:Ce) is a relatively new scintillator material.
Although its effective atomic number of 54 and density of \qty{6.6}{\g/\cubic\cm} make it less efficient than BGO for stopping background particles, it remains a promising material to be used in gamma-ray detectors both as primary detectors and as ACD~\cite{dilillo_etal_2022,yoneyama_etal_2018, HERD_ACS}.
Its light yield is typically in the range of 40 to 60 optical photons per keV~\cite{Kamada2012,Park2023ChemPolish} compared to 8 to 10 optical photons per keV for BGO~\cite{Park2023} making it efficient at lower X-ray/gamma-ray energies.
Compared to scintillators with similar light yields, such as NaI:Tl, GAGG:Ce has the significant advantage that it is non-hygroscopic and that it shows little performance degradation after irradiation~\cite{yoneyama_etal_2018}.

Potential downsides of GAGG:Ce for use as detector on astrophysical missions are the large number of activation lines~\cite{yoneyama_etal_2018}.
While the presence of these activation lines can cause issues for the use of GAGG:Ce as a gamma-ray spectrometer, they are not problematic for usage in an ACD where no spectrometry is performed.
However, the $\gamma$-rays produced by the activation can result a new background component in the primary instrument, which will have to be studied in detail using, for example, Geant4 \cite{Agostinelli2003GEANT4} simulations. Furthermore, GAGG:Ce exhibits a strong afterglow with a duration of the order of days~\cite{dilillo_etal_2022}.
However, for an ACD with a typical energy threshold of 10s of keV the afterglow, which is unlikely to produce several photons within a \si{\micro\s} period, is unlikely to induce false triggers.
For usage in an ACD GAGG:Ce has the advantages of having a low detection threshold, ensured by the large light yield, along with a relatively high stopping power.
Finally, compared to BGO, the light yield and decay time of GAGG:Ce are significantly less temperature dependent.
Previous studies have shown the scintillation decay time to increase by \qty{20}{\percent} when cooling GAGG:Ce from room temperature to \qty{250}{\K}, while the light yield increases only by several percent~\cite{yoneyama_etal_2018}.
Studies of the performance of GAGG:Ce at cryogenic temperatures are however missing prompting the current study. 

This introduction is followed by section~\ref{sec:gagg-properties} describing of the relevant scintillation properties of GAGG:Ce.
Section \ref{sec:GAGG_setup} describes the methods used in this study, including the test setup, the analysis of the measured signals and the calibration of the PMT.
The results of this study are presented in section \ref{sec:GAGG_results} followed by the conclusions in section~\ref{sec:conclusions}.

\section{GAGG Scintillation Properties}\label{sec:gagg-properties}

GAGG:Ce is an inorganic scintillator doped with \ce{Ce^3+}.
The \ce{Ce^3+} introduces localized energy states within the host crystal's band gap, acting as efficient recombination centers.
In this type of scintillator the incoming radiation ionizes the scintillator material producing electron-hole pairs. 
The electrons and holes travel independently within the scintillator until they reach a crystal defect or a recombination site, which in case of GAGG:Ce is one of the \ce{Ce^3+} sites.
Recombination of electrons and holes occur when they are absorbed by the \ce{Ce^3+} ion.
At this point the energy from the electron hole pair is transferred to the ion, thereby exciting its $\textit{4f}$ electron to the $\textit{5d}$ orbital followed by radiative decay back to the $\textit{4f}$ ground state, resulting in the emission of a photon with an energy in the wide \qtyrange[range-phrase={ to }]{450}{800}{\nm} band.
This emission forms the fast component of the scintillation light we observe.
Within GAGG:Ce a slower component arises from energy migration processes within the \ce{Gd^3+} sub-lattice of the host crystal before the energy is transferred to the \ce{Ce^3+} activator~\cite{GAGG_Gd_role}.
At temperatures below \qty{80}{\K} direct emission from \ce{Gd^3+} has also been reported~\cite{GAGG_traps}. 

The number of scintillation photons emitted at time $t$ after an energy deposition can typically be described by two exponential decays,
\begin{equation}\label{eq:1}
    N(t) = A\exp \left(-{\frac {t}{{\tau }_{f}}}\right) + B\exp \left(-{\frac {t}{{\tau }_{s}}}\right),
\end{equation}
where $\tau_f$ and $\tau_s$ are the fast and slow decay time constants, and $A$ and $B$ are the relative amplitudes of the respective emission components.
At room temperature typically $\tau_f \sim \qty{100}{\ns}$ and $\tau_s \sim \qty{500}{\ns}$~\cite{GAGG_fast_slow}.
However, the values reported in the literature vary significantly, in part due to different Ga/Al ratios~\cite{GAGG_Al_ratio}.
Overall, $\tau_f$ has been found to vary in the range of \qtyrange[range-phrase={ to }]{60}{150}{\ns}~\cite{Furuno_2021, GAGG_Al_ratio}.
Higher levels of gallium typically result in lower light yields but faster decay times~\cite{ GAGG_Al_ratio2}, although other studies show no clear signs of an impact on the decay time ~\cite{GAGG_Al_ratio}.
Given the significant variations in the light yield and decay time reported in the literature, two different types of GAGG:Ce were studied here: a high light-yield variant, which was previously used successfully on the GRAPE mission \cite{GRAPE}; and a faster crystal, foreseen to be used on the spectrometer of the POLAR-2 mission \cite{Kole2025UniversalSiPM}.
Unfortunately, the manufacturers declined to provide the exact Ga/Al ratio.

The light yield in GAGG:Ce is affected by thermal quenching.
In this process the energy of the recombined electron-hole pair is used for ionization of \ce{Ce^3+} releasing electrons into the conduction band instead of emitting light.
As a result, optical emission is expected to decrease with increasing temperature.
The presence of traps in the crystal has the opposite effect.
Electrons or holes can be trapped in meta-stable states within potential wells along the way.
The charge carriers may escape these sites by tunneling to the recombination centers directly or through thermal excitations.
The interplay of these two effects results in a peak light yield around \qty{200}{\K} based on studies from ~\cite{GAGG_decay} or \qty{240}{\K} from  ~\cite{yoneyama_etal_2018}.

Furthermore, the trapping of charge carriers results in an afterglow, which is particularly bright in case of GAGG:Ce~\cite{dilillo_etal_2022}.
The release of the trapped charges will be shorter at high temperatures due to thermal excitations, while at lower temperatures trapped charges can remain stuck until the crystal is warmed up again~\cite{GAGG_traps}.
As a result, when the temperature is lowered, a larger fraction of charges will become permanently trapped, possibly until the temperature is increased.
Therefore, fewer charge carriers will reach the activation centers.
the afterglow intensity therefore increases at higher temperatures~\cite{yoneyama_etal_2018}.
Furthermore, \citep{GAGG_decay} finds that the decay time increases with decreasing temperature, down to around \qty{\sim 200}{\K}. Below this temperature, the permanent trapping of charges causes delayed emission to decrease, thereby reducing the light yield, and decreasing the decay time. While other complex effects in the crystals can also contribute to the temperature dependence.

\section{Methods}
\subsection{Experimental Setup}\label{sec:GAGG_setup}

\begin{figure}
  \centering
  \includegraphics[width=0.7\textwidth]{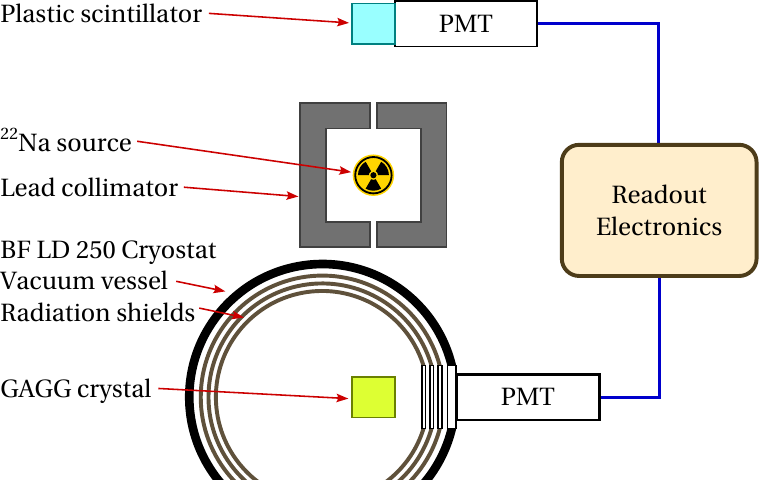}
  \caption{%
    Schematic of the setup used to measure the performance of GAGG at different temperatures.
    The GAGG sample is connected to the mixing chamber stage of a BlueFors LD 250 cryostat.
    Optical windows in the radiation shields and the vacuum vessel allow observing the GAGG from the outside using a photomultiplier tube coupled to the vacuum window with a light-tight seal.
    The GAGG is illuminated by a \qty{\sim 10.8}{\mega \mathrm{Bq}} (\qty{\sim 290}{\micro\curie}) $^{22}\text{Na}$ source placed at approximately \qty{0.7}{\m} distance in a collimated lead enclosure outside the cryostat.
    The isotope $^{22}\text{Na}$ is a positron emitter resulting in the emission of two back-to-back \qty{511}{\keV} gamma-rays.
    A fast plastic scintillator coupled directly to a PMT is placed equidistantly opposite the GAGG.
    The signal from both PMTs is used to create a coincidence signal, which triggers the readout of signals from the PMT observing the GAGG as shown in figure~\ref{fig:GAGG_electronics}.%
  }
  \label{fig:GAGG_setup}
\end{figure}

In order to study the effect of temperature on the GAGG:Ce, independent of thermal effects on the scintillation light detector, a setup was developed where only the temperature of the GAGG crystal is varied similar to the method employed in~\cite{PhysRevB.84.214306}.
In this setup, shown schematically in figure~\ref{fig:GAGG_setup}, the GAGG:Ce crystal is placed in a cryostat (Bluefors LD 250) where it is thermally coupled to the final stage which can be cooled down to \qty{15}{\milli\K}. 
The crystal is visible from outside the cryostat through windows placed in the 4 shells of the cryostat (vacuum vessel, 50K stage, 4K stage and the still heat shield).
A Hamamatsu R1924A PMT~\cite{R1924A} is placed on the outside of the vacuum vessel looking through the outer window. The PMT has a good thermal connection to the vacuum vessel, which in turn has a temperature equal to the room temperature of the laboratory. This temperature, and therefore the temperature of the PMT has been measured to be stable within approximately 1 K, over a period of several days.

\begin{figure}
  \centering
  \includegraphics[width=1.0\textwidth]{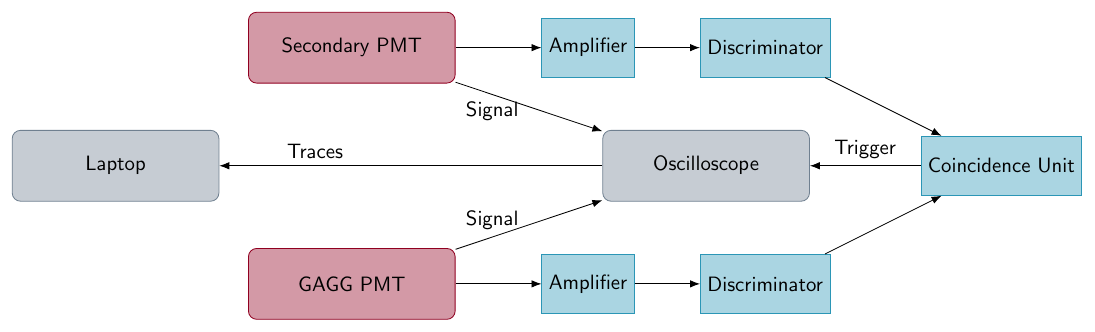}
  \caption{%
    Schematic of the readout electronics from figure \ref{fig:GAGG_setup} used for measuring the relative light yield and the decay time of the GAGG scintillators.
    Signals from the two PMTs are sent to amplifiers.
    The amplified signal is processed by discriminators whose output is passed to a coincidence unit.
    In case of a coincidence a trigger is sent to an oscilloscope which reads out the non-amplified signals of the two PMTs.
    The traces from both PMTs are read out using a laptop computer and stored for further analysis.%
  }
  \label{fig:GAGG_electronics}
\end{figure}

The disadvantage of this setup is that the fraction of optical photons that reach the PMT is small.
Optical simulations performed using Geant4 \cite{Agostinelli2003GEANT4} indicate that at room temperature a \qty{511}{keV} deposition in the GAGG crystal results in approximately 8 optical photons reaching the PMT on average.
Given the quantum efficiency of the PMT of \qty{\sim 25}{\percent} only an average of 2 or 3 photoelectrons are expected per gamma-ray.
The fact that the photons will arrive within a period of approximately \qty{100}{\ns} makes detecting this signal challenging.
To distinguish the single photoelectron signals from noise, a coincidence setup which employs a second PMT of the same type is used. The PMT is connected to an EJ-200 plastic scintillator which has a rise time of \qty{0.9}{\ns} and a decay time of \qty{2.1}{\ns} \cite{Eljen}. Both PMTs are equipped with a preamplifier, originally developed for the NSPECT instrument~\cite{ryan_nspect_2010}, which provides an amplification of $\sim 10$ while the rise time remains within the order of~\qty{1}{\ns}. 

A $^{22}\text{Na}$ source was placed in a collimated lead enclosure between the GAGG:Ce and the plastic scintillator. 
This allows the two scintillators to detect the coincident back-to-back \qty{511}{\keV} photons emitted from this source.
The output from both PMTs is split with one signal going directly to an oscilloscope, while the second is passed to spectral amplifiers in a NIM crate as indicated in Fig.~\ref{fig:GAGG_electronics}.
The output of the amplifiers, with an amplification of 100 and a shaping time of \qty{250}{\ns}, is passed to discriminators.
A logic unit is then used to produce a coincidence signal if the outputs from both discriminators are high within a \qty{1}{\micro\s} window.
The output signal from the logic unit triggers the readout of an oscilloscope which records the two non-amplified output signals from the PMTs.
Traces for both signals were recorded over a \qty{5}{\micro\s} period with a sampling rate of \qty{0.5}{\GHz}.
This setup provides an accurate trigger time and allows to efficiently record the signals from the GAGG:Ce, which consist of only a few single photoelectron pulses spread over a relatively long period.

\subsection{Pulse Analysis}\label{sec:PA}

\begin{figure}
  \centering
  \includegraphics[width=0.7\textwidth]{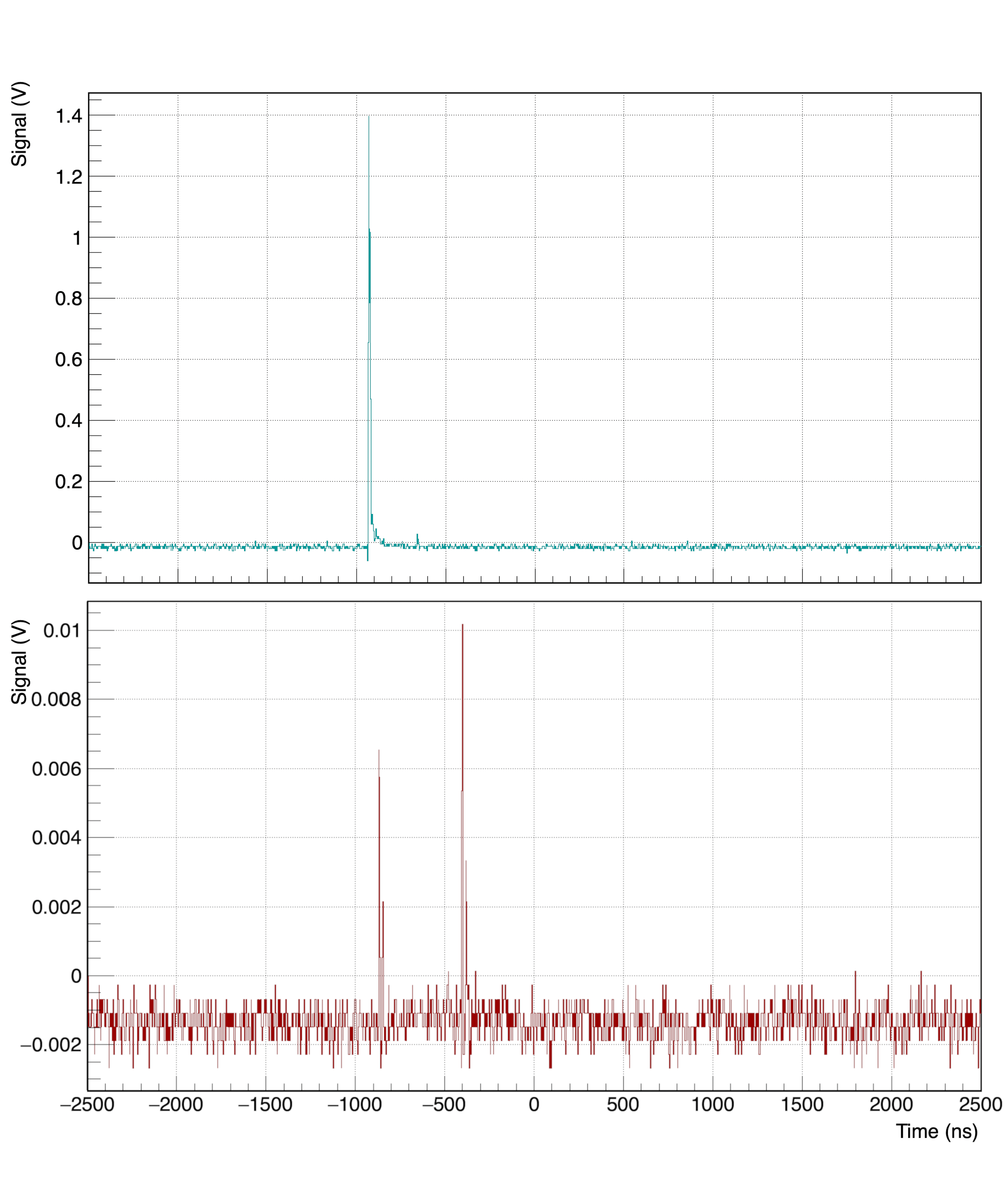}
  \caption{%
    An example of the traces from the two PMTs for a valid coincident event.
    The top figure shows the trace from the PMT connected directly to a plastic scintillator, while the bottom one shows the signal from the PMT observing the GAGG:Ce crystal.
    The signal from the GAGG:Ce consists of 2 pulses from 2 separate photo-electrons.
    A consistent reflection from the signal in the cables can be seen for each pulse which is cleaned up in the analysis.%
  }
  \label{fig:trace_example}
\end{figure}

\begin{figure}
  \centering
  \includegraphics[width=0.7\textwidth]{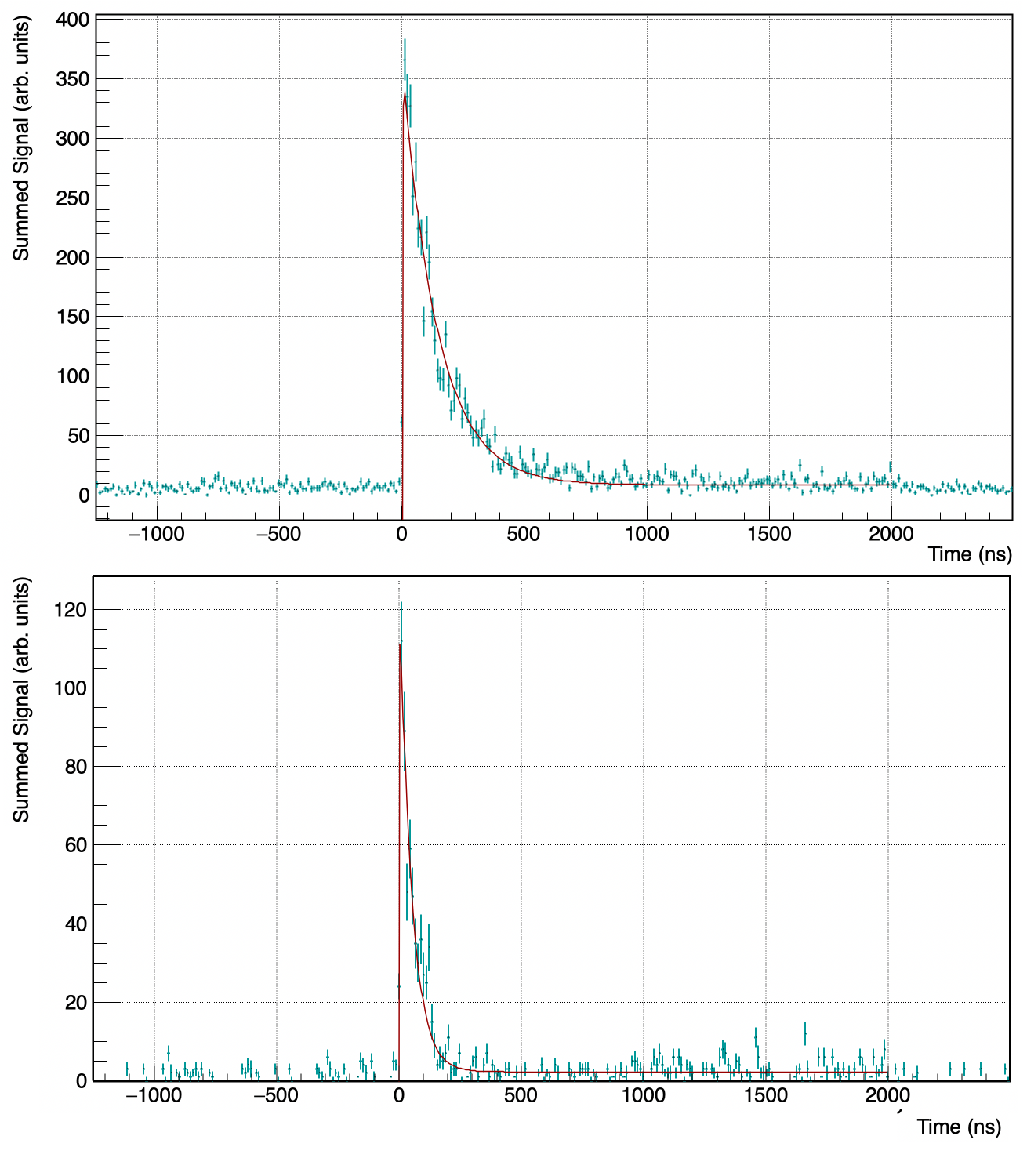}
  \caption{%
    \textbf{Top:} Example of the reconstructed scintillation pulses produced by summing 2400 traces taken at room temperature from the first GAGG:Ce crystal after aligning them based on the peak position of the pulse in the plastic scintillator.
    The pulse is fitted with the function given in equation~\eqref{eq:2}.
    \textbf{Bottom:} Same as the top figure but for 800 traces taken at \qty{50}{\milli\K}.%
  }
  \label{fig:pulse_examples}
\end{figure}

A typical example of a recorded set of traces for a coincident event are shown in Fig.~\ref{fig:trace_example}.
The traces will have to be corrected for time walk effects as the trigger arrival time on the oscilloscope is produced through the amplified signal. Due to shaping of the amplifiers this is significantly longer. As a result, the moment the signal passes the discriminator depends on the signal height and therefore the time at which the oscilloscope holds the two PMT signals varies.
To mitigate this in the analysis, the peak time of the plastic PMT signal is found and used to shift the trace from the GAGG:Ce PMT such that $t=0$ corresponds to the peak of the plastic scintillator signal.
Using this method, the GAGG:Ce pulses, which typically consist of a few single photoelectron signals, can be alligned and then summed to produce an average pulse signal from \qty{511}{\keV} photons.
Examples of resulting summed pulses are shown in figure~\ref{fig:pulse_examples}, one resulting from 800 traces recorded with the GAGG at \qty{50}{\milli\K} and one from 2400 traces recorded at room temperature.

Data were taken at a number of temperatures between \qty{15}{\milli\K} and room temperature. The cryostat was first cooled down to this temperature, which took approximately 2 days. In the range of \qty{15}{\milli\K} up to around \qty{4}{\K} heaters in the cryostat were subsequently used to stabilize the system at various temperatures in this range. At higher temperatures stabilization was not possible.
As a result, detailed data at fixed temperatures could be taken at room temperature and below \qty{4}{\K}, while data at intermediate temperatures was acquired while the cryostat was warming up.
Although the temperature of the GAGG:Ce was therefore not stable, the slow warming of the cryostat (approximately 1.5 days) allowed us to record sets of traces within several Kelvin above \qty{4}{\K}.
Between \qty{4}{\K} and \qty{40}{\K}, the warm up proceeded relatively fast resulting in a systematic error on the temperature of several degrees.
Above  \qty{40}{\K} it becomes possible to acquire enough data within a \qty{5}{\K} temperature interval to produce significant measurements.
However, because the main interest of this study is to measure the performance at \qty{4}{\K} and below, the data at higher temperatures are mainly used to validate our measurements against previously published results.

\subsection{PMT Calibration}\label{sec:PMT_calib}

The pulses produced by summing the traces can, in principle, be used to give an indication of both the decay time and the relative pulse height.
However, given the weak signal, and to ensure we account correctly for the Poisson statistics of the number of detected photoelectrons, the average number of detected photoelectrons in each pulse was used instead.
This number is proportional to the light yield and its distribution should be Poissonian.

In order to count the number of photoelectrons in a trace, we first need to know the mean pulse height per photoelectron.
This required calibration of the PMT and its readout logic, which was achieved by injecting light pulses of \qty{50}{\ns} duration into the cryostat using an optical fiber.
This test was performed at room temperature but with an otherwise unchanged setup in order to avoid any differences in light collection.
%
%
A low intensity of the light source was selected such that only several photoelectrons were observed by the PMT.
The signal was again 
read out using an oscilloscope, which in this case was triggered using a signal from the light source.

\begin{figure}
  \centering
  \includegraphics[width=1.0\textwidth]{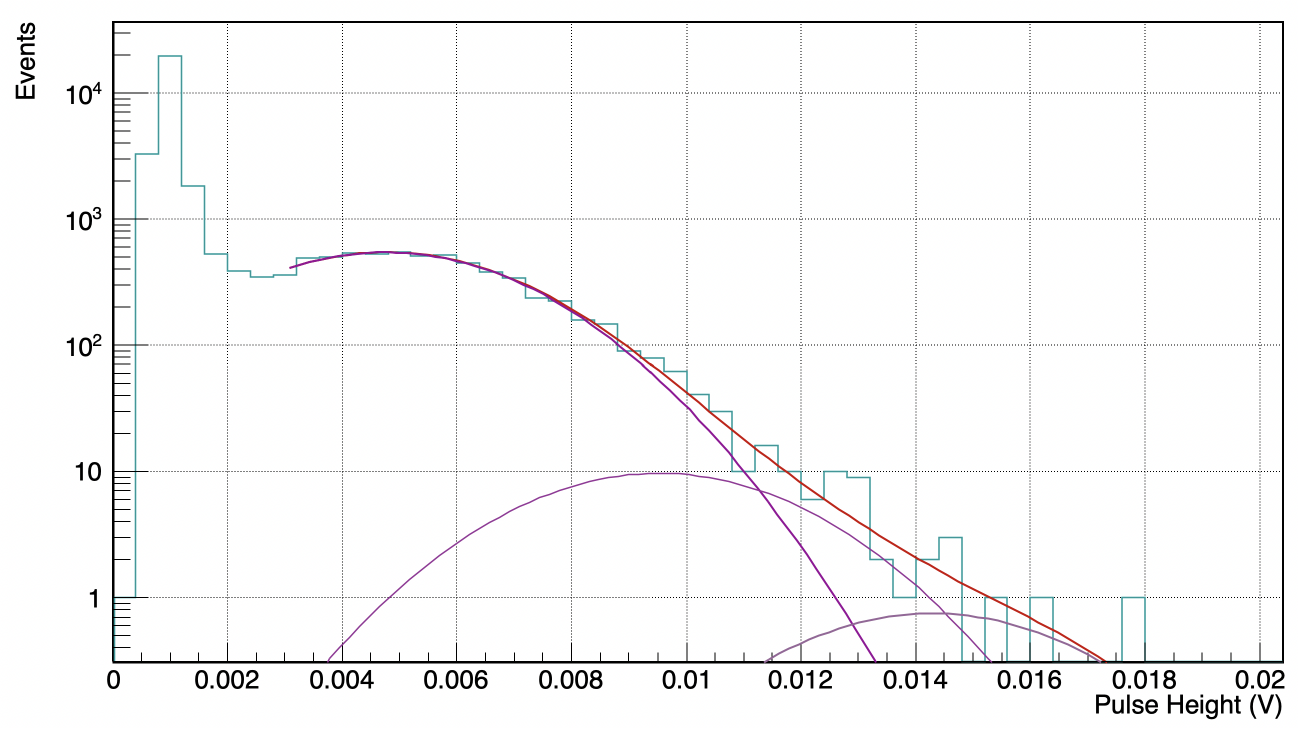}
  \caption{%
    The pulse height spectrum from the PMT acquired through illumination with a pulsed light source. The dynode peak can be seen at around \qty{1}{\mV} followed by a continuum produced by the sum of the  photoelectron peaks.%
  }
  \label{fig:Low_light}
\end{figure}

\begin{figure}
  \centering
  \includegraphics[width=1.0\textwidth]{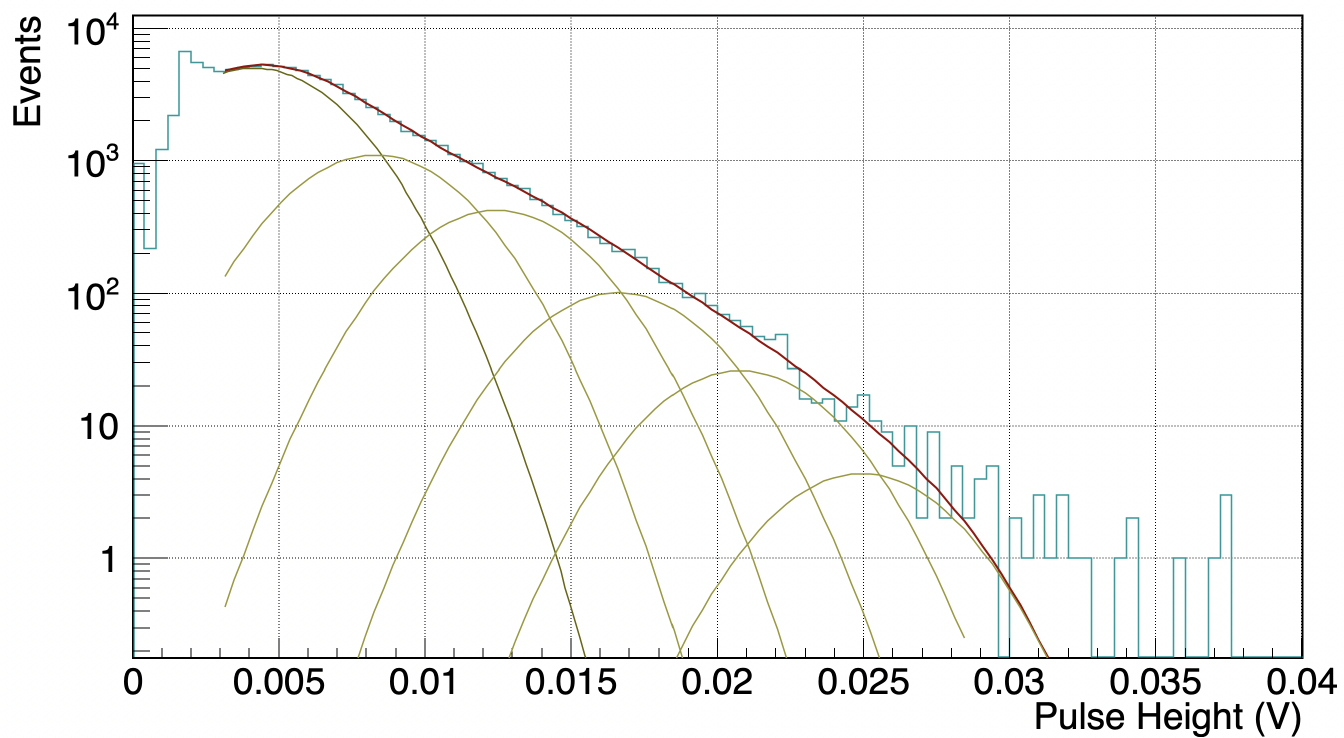}
  \caption{%
    Similar to figure \ref{fig:Low_light} but with the light source set 8 times higher intensity.%
  }
  \label{fig:More_light}
\end{figure}

A total of \num{10000} traces were recorded at several different light intensities.
Examples of the measured pulse height spectrum for both the brightest and weakest pulse intensity are shown in figures~\ref{fig:Low_light} and~\ref{fig:More_light}.
The spectra show a peak at around \qty{1}{\mV} which can be attributed to the 'first dynode effect'~\cite{Tokar1999}.
In such events a photon is not absorbed in the photocathode but instead causes a photoelectron to be emitted from the first dynode.
The continuum after this can be well fitted using a sum of equidistant normal distributions.
In the fit, the width of the standard deviation ($\sigma$) of all the distributions is forced to be equal.
The distance between the peaks is equal to the mean of the first peak.
Therefore, the free parameters of the fit are $\sigma$, the mean of the first peak, and the amplitudes of the different peaks.
As expected, increasing the light intensity increases the average number of observed photoelectrons, 
while pulse height per photo-electron does not change. 

\begin{figure}
  \centering
  \includegraphics[width=1.0\textwidth]{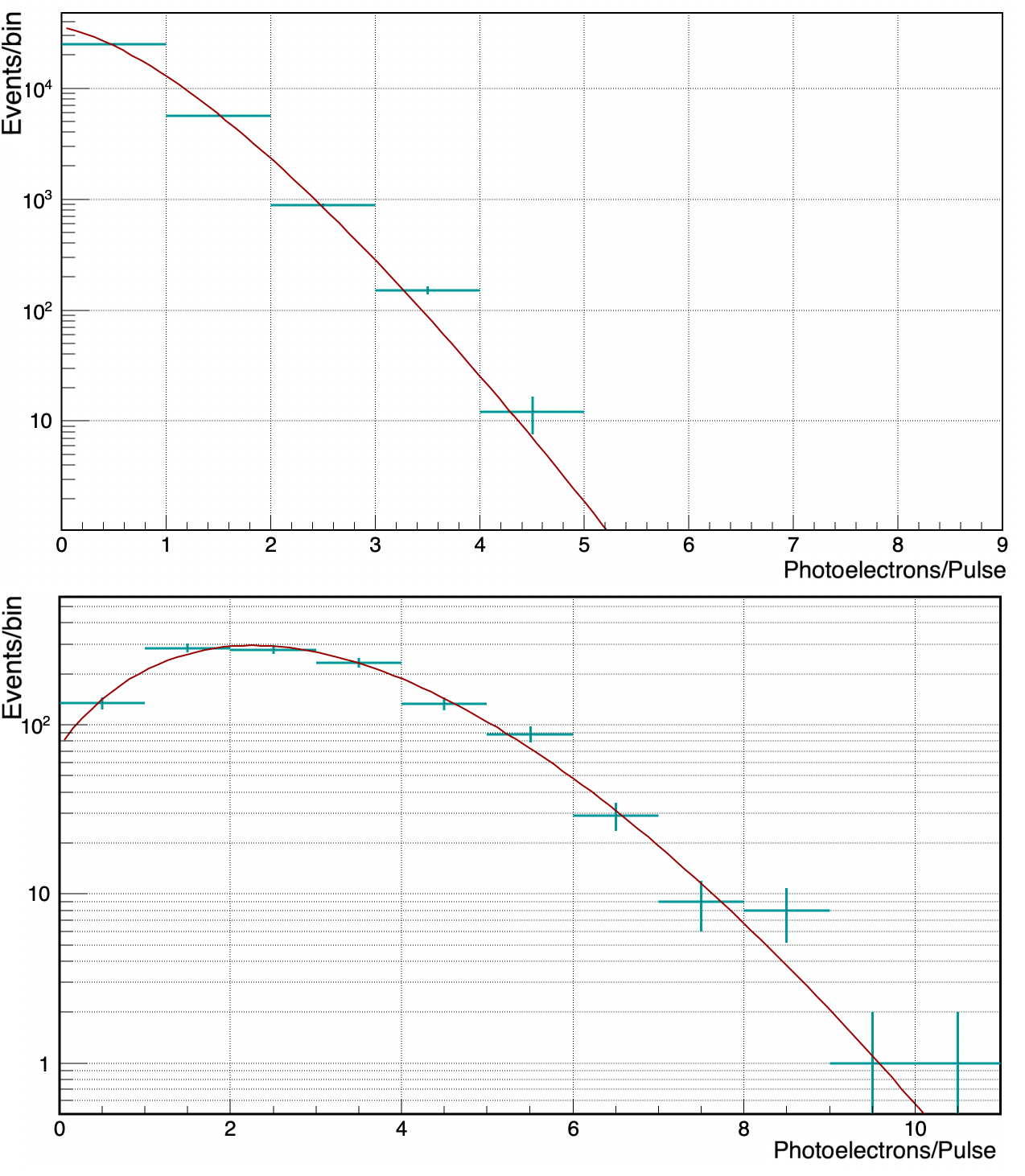}
  \caption{%
    The distribution of the number of detected photons from the pulsed light source for the weakest light pulses (top) and strongest light pulses (bottom) used in this calibration.%
  }
  \label{fig:poisson}
\end{figure}

By fitting the spectra in \ref{fig:More_light} and \ref{fig:Low_light} in this way we determined a mean amplitude per photoelectron of \qty{4.4+-0.2}{\mV} and \qty{4.3+-0.2}{\mV}, respectively.
We therefore used a value of \qty{4.35}{\mV}  to count the number of photoelectrons in the traces produced by the light source.
To count the number of photoelectrons in a trace, the height of each pulse was divided by the mean pulse height and rounded up to the nearest integer.
The distribution of the number of photoelectrons from both the weak and the strong light pulses are shown in figure~\ref{fig:poisson}.
Both distributions follow the expected Poisson distribution with which they are fitted.
The fits find a mean number of detected photoelectrons of $0.36\,\mathrm{p.e./pulse}$ with the weak light pulses and $2.8\,\mathrm{p.e./pulse}$ for the stronger light pulses.

\section{Results}\label{sec:GAGG_results}

\begin{figure}
  \centering
  \includegraphics[height=0.4\textwidth]{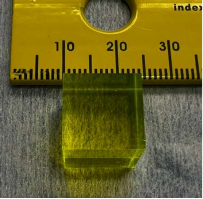}
    \includegraphics[height=0.4\textwidth]{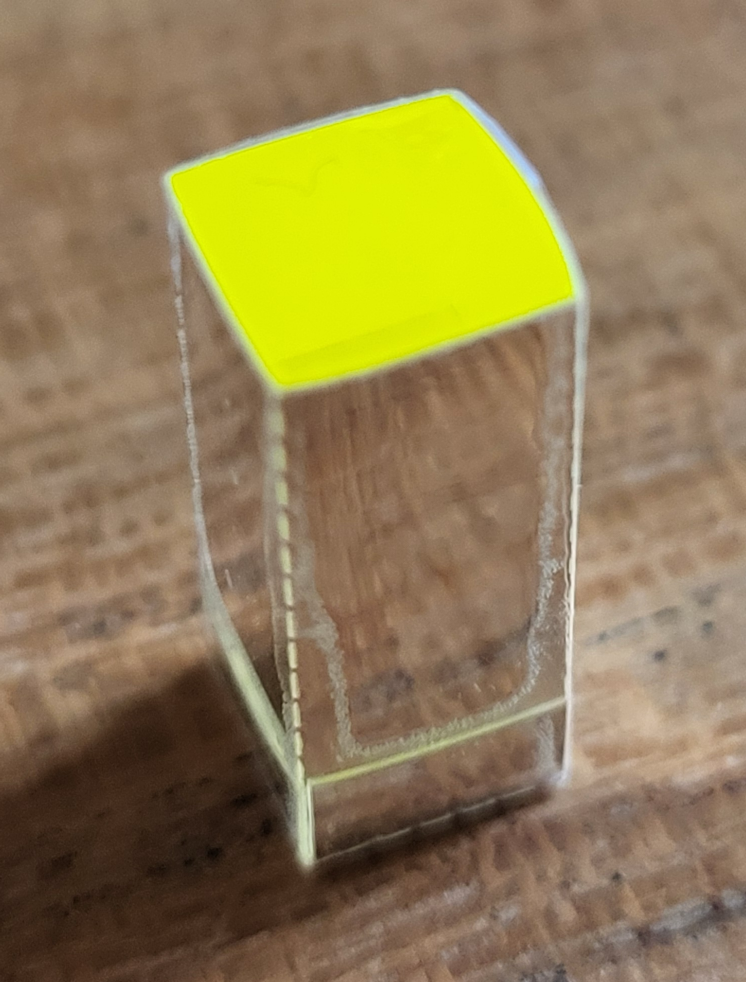}

  \caption{%
    The first crystal, previously used in the GRAPE mission, used for this test on the left and the second crystal, wrapped in reflective material on the sides, on the right. %
  }
  \label{fig:crystals}
\end{figure}

The crystal tested first concerns a \qtyproduct{12 x 12 x 12}{\mm} ``balanced'' GAGG:Ce scintillator, shown on the left in Figure \ref{fig:crystals}, from Hangzhou Yong Hee Photonics previously used on the GRAPE balloon mission~\cite{GRAPE}.
Here, ``balanced'' refers to the aluminum to gadolinium ratio chosen to provide a balance between high light yield and fast decay time
~\cite{yphotonics154}.
The second tested crystal, shown on the right in Figure \ref{fig:crystals}, has dimensions of \qtyproduct{6 x 6 x 20}{\mm} and was provided by the POLAR-2 collaboration \cite{Kole2025UniversalSiPM}.
It is faster at room temperature compared to the first crystal.

\subsection{Decay Time}

At each temperature, at least 600 traces were recorded and summed to produce averaged pulses such as those presented in Fig.~\ref{fig:pulse_examples}.
These were initially fitted using equation~\eqref{eq:1}.
However, the slow component was typically not well constrained.
Using data from a 3 hour measurement at room temperature the slow component was constrained to have a decay time of \qty{\sim800}{\ns} but was only responsible for \qty{\sim 3}{\percent} of the emission.
We therefore simplified the analysis and fitted all other data using a single decay component similar to Ref.~\cite{yoneyama_etal_2018}.
In addition, to account for potential effects from the rise time of the signal and the shift of the pulse peak position, all results were fitted using:
\begin{equation}\label{eq:2}
    N(t) = A\left[\exp \left(-{\frac {t-t_\mathit{peak}}{{\tau_\mathit{decay} }}}\right) - \exp \left(-{\frac {t-t_\mathit{peak}}{\tau_\mathit{rise}}}\right)\right]+B.
\end{equation}
Here $\tau_\mathit{decay}$ is the decay time of the pulse $\tau_\mathit{rise}$ the rise time of the signal, $t_\mathit{peak}$ the position of the pulse peak, $A$ the amplitude of the pulse, and $B$ a parameter to account for a potential offset of the signal due to chance coincidences which make up a few percent of the data.
The fit was performed in the range $t=0$ to $t=\qty{2000}{\ns}$.

\begin{figure}
  \centering
  \includegraphics[width=1.0\textwidth]{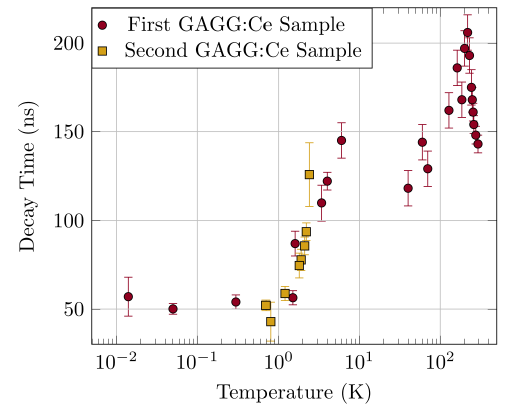}
  \caption{%
    The decay time of the GAGG:Ce crystal from GRAPE as a function of temperature. The round data points were taken with a first GAGG:Ce crystal while the square points were taken with a second sample of the same type to check for consistency.%
  }
  \label{fig:GAGG_decay}
\end{figure}

The results for the crystals from the GRAPE mission are summarized in figure \ref{fig:GAGG_decay}. 
When the temperature is decreased the pulses initially become longer, peaking at around \qty{200}{\K}, after which they become shorter again.
It should be noted again that at these high temperatures the measurements were performed while the system was heating up.
Therefore the crystal will not be at the same temperature as the measurement points in the cryostat.
Therefore, a systematic error of \qty{\sim5}{\K} on the temperature exists above \qty{4}{\K}. Although, as a result, we cannot make strong claims on the behavior at high temperatures, the results match previous measurements in this range \cite{yoneyama_etal_2018}.
In addition, we observe an abrupt change at \qty{\sim 2}{\K} where the decay time drops from \qty{\sim130}{\ns} to \qty{\sim50}{\ns}.
Below this drop, the decay time again remains relatively stable. 

We can conclude firstly that at \qty{4}{\K} the decay time is similar to that at room temperature.
Therefore, this will not cause any issues for usage in an ACD at this temperature.
In addition, we observe a sharp drop at \qty{\sim2}{\K}.
A measurement with a second crystal of the same type was performed to confirm this (square markers in figure \ref{fig:GAGG_decay}).
The results show a consistent behavior, while the additional measurement points indicate that the transition from \qty{130}{\ns} to \qty{\sim50}{\ns} is sharp and occurs at around \qty{2}{\K}.
Measurements at \qty{2.1}{\K} and above consistently produced decay times consistent with \qty{130}{\ns}, while all those of \qty{1.8}{\K} and below show decay times below \qty{70}{\ns}.

\begin{figure}
  \centering
  \includegraphics[width=1.0\textwidth]{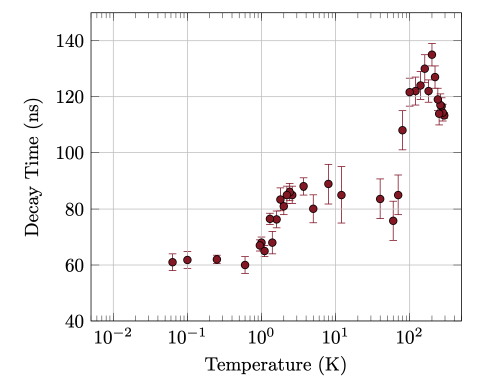}
  \caption{%
    The decay time of the GAGG:Ce crystal from POLAR-2 as a function of temperature.%
  }
  \label{fig:IHEP_crystal_decay}
\end{figure}

The results for the GAGG:Ce crystal from POLAR-2 are summarized in figure \ref{fig:IHEP_crystal_decay}.
The overall characteristics of the temperature dependence are very similar to that seen for the GRAPE crystal in \ref{fig:GAGG_decay}.
At room temperature this type is faster with a decay time of \qty{113}{\ns} compared to \qty{138}{\ns}.
Similar to the GRAPE crystal, the decay time initially increases reaching a peak around \qty{200}{\K}, followed by a decrease before plateauing around \qty{70}{\K}.
Again a sharp drop is seen at around \qty{2}{\K}, below which, similar to the other crystal the decay time is \qty{\sim60}{\ns}.
Therefore, both types of crystal show a similar decay time only below \qty{2}{\K}, whereas at higher temperature the general features are similar but decay times differ significantly.

Finally, we note that although $\tau_\mathit{rise}$ was included in all fits, the rising edge was consistently found to be within 1 or 2 samples.
Therefore, no significant measurements of this parameters were possible but we conclude that the rise time stays within the order of several ns throughout the full temperature range.

\subsection{Light Yield}

\begin{figure}
  \centering
  \includegraphics[width=1.0\textwidth]{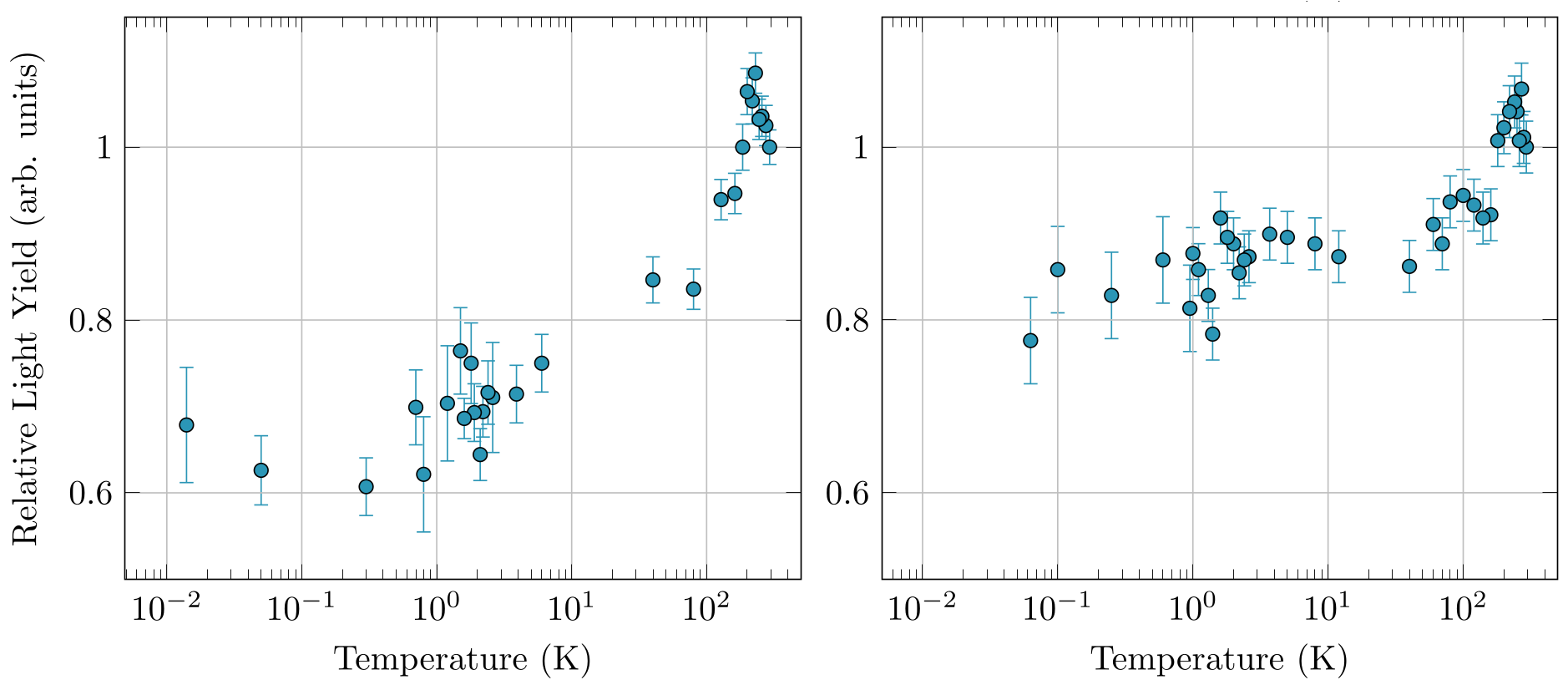}
  \caption{%
    The relative number of mean observed photoelectrons produced by interactions in the the GAGG:Ce crystal as a function of temperature both for both the crystal from GRAPE (left) and from POLAR-2 (right).%
  }
  \label{fig:GAGG_light}
\end{figure}

Using the same analysis as applied for the PMT calibration, we can count the number of photoelectrons produced from the scintillation pulses from the GAGG:Ce crystal.
For each temperature measurement the number of pulses in the trace are counted and their height is divided by the mean pulse height value of \qty{4.35}{\mV}, followed by rounding off to the nearest integer. 
The number of photoelectrons per trace are then entered in a histogram which is fitted by a Poisson distribution.
These distributions were consistently well fitted with reduced $\chi^2$ values in the 0.8 to 1.8 range.
The Poisson mean is taken as the light yield and plotted as a function of temperature for both crystals in figure \ref{fig:GAGG_light}. 

When decreasing the temperature from room temperature the light yield increases before reaching a maximum around \qty{220}{\K}, matching the results found in~\cite{yoneyama_etal_2018}.
Below the peak, the light yield decreases before stabilizing around \qty{70}{\percent} of the light yield at room temperature. The POLAR-2 crystal shows a very similar behavior, although the plateau here is around $80\%$.
Whereas there is a clear change in the decay time at \qty{2}{\K}, the light yield does not significantly change at this temperature.
Comparing the two crystals the structure again looks very similar although the variations in the second crystal with temperature are smaller.

\section{Conclusions}\label{sec:conclusions}

We measured both the relative light yield and the decay time of two types of GAGG:Ce between room temperature and cryogenic temperatures for the first time.
The results indicate that both parameters change relatively little within this range.
The light yield at \qty{4}{\K}, where we aim to operate such crystals in an ACD, is, for the crystal from GRAPE, approximately \qty{30}{\percent} lower than at room temperature. For the crystal from POLAR-2 it is about \qty{20}{\percent} lower.
Therefore, the decrease in light yield is unlikely to hamper the use of such a material in a cryogenic ACD.
The decay time of the scintillation signal at \qty{4}{\K} was found to be similar to room temperature.
For the crystal as used in GRAPE it is within \qty{10}{\percent} of that at room temperature and for the one from POLAR-2 it is \qty{30}{\percent} faster.

Of further interest is the significant decrease of the decay time observed at \qty{2}{\K} for both crystals.
For both types the drop occurs at the same temperature and in both cases the decay time decreases to \qty{\sim 60}{\ns}.
However, the decay times for the full temperature range above \qty{2}{\K} are significantly different for the two types.
The GRAPE has a decay time of \qty{\sim 130}{\ns} between \qty{3}{\K} and \qty{80}{\K} while for the crystal from POLAR-2 this is \qty{\sim 90}{\ns}.
Overall, the region below \qty{2}{\K} is the only temperature interval where the decay times of the two types are similar.
No significant changes in light yield were observed between \qty{1}{\K} and \qty{3}{\K}, indicating that the same amount of light is emitted in approximately half the time when GAGG:Ce is cooled below \qty{2}{\K}. 
Although this effect does not affect the use of GAGG:Ce for our purposes, we welcome suggestions on the cause of this phenomenon. Similar effects have, to our knowledge, not been found in other scintillators at such temperatures. It is, however, known that changes to the crystal structure do occur in other gadolinium garnets at cryogenic temperatures. Studies presented in \cite{GAG} show sharp changes to the specific heat to occur at such temperatures for $\mathrm{Gd_3Al_5O_{12}}$. A future interdisciplinary study could allow to understand whether such changes are related.

\bmhead{Acknowledgements}
We thank Prof.\ Jianchao Sun from the Institute of High Energy Physics (IHEP) for providing the GAGG:Ce sample from the POLAR-2 mission. This work is supported in part by NASA Astrophysics Research and Analysis grant 80NSSC22K1891.
TD is grateful for support from an Undergraduate Research Award by the University of New Hampshire Hamel Center for Undergraduate Research.

\section*{Declarations}
\subsection*{Funding}
This work is supported in part by NASA Astrophysics Research and Analysis grant 80NSSC22K1891.
TD received support in the form of an Undergraduate Research Award from the University of New Hampshire Hamel Center for Undergraduate Research.

\subsection*{Competing interests}
The authors have no financial or proprietary interests in any material discussed in this article.

\subsection*{Ethics approval and consent to participate}
Not applicable.

\subsection*{Consent for publication}
Not applicable.

\subsection*{Data availability}
All data supporting the findings of this study are available from Zenodo archive https://doi.org/10.5281/zenodo.17412779.

\subsection*{Code availability}
All code used in the analysis of data presented in this paper is available from Zenodo archive https://doi.org/10.5281/zenodo.17413013.

\subsection*{Author contribution}
F.K.\ conceived the experiments, which were carried out by M.K., K.W., R.G., T.D., and Z.G.; M.K.\ designed and carried out the data analysis; M.K.\ wrote the paper with contributions from F.K.


\bibliography{gagg-4k}


\begin{thebibliography}{40}
\ifx \bisbn   \undefined \def \bisbn  #1{ISBN #1}\fi
\ifx \binits  \undefined \def \binits#1{#1}\fi
\ifx \bauthor  \undefined \def \bauthor#1{#1}\fi
\ifx \batitle  \undefined \def \batitle#1{#1}\fi
\ifx \bjtitle  \undefined \def \bjtitle#1{#1}\fi
\ifx \bvolume  \undefined \def \bvolume#1{\textbf{#1}}\fi
\ifx \byear  \undefined \def \byear#1{#1}\fi
\ifx \bissue  \undefined \def \bissue#1{#1}\fi
\ifx \bfpage  \undefined \def \bfpage#1{#1}\fi
\ifx \blpage  \undefined \def \blpage #1{#1}\fi
\ifx \burl  \undefined \def \burl#1{\textsf{#1}}\fi
\ifx \doiurl  \undefined \def \doiurl#1{\url{https://doi.org/#1}}\fi
\ifx \betal  \undefined \def \betal{\textit{et al.}}\fi
\ifx \binstitute  \undefined \def \binstitute#1{#1}\fi
\ifx \binstitutionaled  \undefined \def \binstitutionaled#1{#1}\fi
\ifx \bctitle  \undefined \def \bctitle#1{#1}\fi
\ifx \beditor  \undefined \def \beditor#1{#1}\fi
\ifx \bpublisher  \undefined \def \bpublisher#1{#1}\fi
\ifx \bbtitle  \undefined \def \bbtitle#1{#1}\fi
\ifx \bedition  \undefined \def \bedition#1{#1}\fi
\ifx \bseriesno  \undefined \def \bseriesno#1{#1}\fi
\ifx \blocation  \undefined \def \blocation#1{#1}\fi
\ifx \bsertitle  \undefined \def \bsertitle#1{#1}\fi
\ifx \bsnm \undefined \def \bsnm#1{#1}\fi
\ifx \bsuffix \undefined \def \bsuffix#1{#1}\fi
\ifx \bparticle \undefined \def \bparticle#1{#1}\fi
\ifx \barticle \undefined \def \barticle#1{#1}\fi
\bibcommenthead
\ifx \bconfdate \undefined \def \bconfdate #1{#1}\fi
\ifx \botherref \undefined \def \botherref #1{#1}\fi
\ifx \url \undefined \def \url#1{\textsf{#1}}\fi
\ifx \bchapter \undefined \def \bchapter#1{#1}\fi
\ifx \bbook \undefined \def \bbook#1{#1}\fi
\ifx \bcomment \undefined \def \bcomment#1{#1}\fi
\ifx \oauthor \undefined \def \oauthor#1{#1}\fi
\ifx \citeauthoryear \undefined \def \citeauthoryear#1{#1}\fi
\ifx \endbibitem  \undefined \def \endbibitem {}\fi
\ifx \bconflocation  \undefined \def \bconflocation#1{#1}\fi
\ifx \arxivurl  \undefined \def \arxivurl#1{\textsf{#1}}\fi
\csname PreBibitemsHook\endcsname

\bibitem[\protect\citeauthoryear{Wollman et~al.}{2000}]{WOLLMAN2000145}
\begin{barticle}
\bauthor{\bsnm{Wollman}, \binits{D.A.}},
\bauthor{\bsnm{Nam}, \binits{S.W.}},
\bauthor{\bsnm{Newbury}, \binits{D.E.}},
\bauthor{\bsnm{Hilton}, \binits{G.C.}},
\bauthor{\bsnm{Irwin}, \binits{K.D.}},
\bauthor{\bsnm{Bergren}, \binits{N.F.}},
\bauthor{\bsnm{Deiker}, \binits{S.}},
\bauthor{\bsnm{Rudman}, \binits{D.A.}},
\bauthor{\bsnm{Martinis}, \binits{J.M.}}:
\batitle{Superconducting transition-edge-microcalorimeter x-ray spectrometer with 2ev energy resolution at 1.5kev}.
\bjtitle{Nucl. Instrum. Meth. A}
\bvolume{444}(\bissue{1}),
\bfpage{145}--\blpage{150}
(\byear{2000})
\doiurl{10.1016/S0168-9002(99)01351-0}
\end{barticle}
\endbibitem

\bibitem[\protect\citeauthoryear{{Gottardi} and {Nagayoshi}}{2022}]{2022arXiv221006914G}
\begin{botherref}
\oauthor{\bsnm{{Gottardi}}, \binits{L.}},
\oauthor{\bsnm{{Nagayoshi}}, \binits{K.}}:
{A Review of X-ray Microcalorimeters Based on Superconducting Transition Edge Sensors for Astrophysics and Particle Physics}.
arXiv e-prints,
2210--06914
(2022)
\doiurl{10.48550/arXiv.2210.06914}
{\href{https://arxiv.org/abs/2210.06914}{{arXiv:2210.06914}}}
{[astro-ph.IM]}
\end{botherref}
\endbibitem

\bibitem[\protect\citeauthoryear{Barret et~al.}{2023}]{Barret2023}
\begin{barticle}
\bauthor{\bsnm{Barret}, \binits{D.}},
\bauthor{\bsnm{Albouys}, \binits{V.}},
\bauthor{\bsnm{Herder}, \binits{J.-W.}},
\bauthor{\bsnm{Piro}, \binits{L.}},
\bauthor{\bsnm{Cappi}, \binits{M.}},
\bauthor{\bsnm{Huovelin}, \binits{J.}},
\bauthor{\bsnm{Kelley}, \binits{R.}},
\bauthor{\bsnm{Mas-Hesse}, \binits{J.M.}},
\bauthor{\bsnm{Paltani}, \binits{S.}},
\bauthor{\bsnm{Rauw}, \binits{G.}},
\bauthor{\bsnm{Rozanska}, \binits{A.}},
\bauthor{\bsnm{Svoboda}, \binits{J.}},
\bauthor{\bsnm{Wilms}, \binits{J.}},
\bauthor{\bsnm{Yamasaki}, \binits{N.}},
\bauthor{\bsnm{Audard}, \binits{M.}},
\bauthor{\bsnm{Bandler}, \binits{S.}},
\bauthor{\bsnm{Barbera}, \binits{M.}},
\bauthor{\bsnm{Barcons}, \binits{X.}},
\bauthor{\bsnm{Bozzo}, \binits{E.}},
\bauthor{\bsnm{Ceballos}, \binits{M.T.}},
\bauthor{\bsnm{Charles}, \binits{I.}},
\bauthor{\bsnm{Costantini}, \binits{E.}},
\bauthor{\bsnm{Dauser}, \binits{T.}},
\bauthor{\bsnm{Decourchelle}, \binits{A.}},
\bauthor{\bsnm{Duband}, \binits{L.}},
\bauthor{\bsnm{Duval}, \binits{J.-M.}},
\bauthor{\bsnm{Fiore}, \binits{F.}},
\bauthor{\bsnm{Gatti}, \binits{F.}},
\bauthor{\bsnm{Goldwurm}, \binits{A.}},
\bauthor{\bsnm{Hartog}, \binits{R.}},
\bauthor{\bsnm{Jackson}, \binits{B.}},
\bauthor{\bsnm{Jonker}, \binits{P.}},
\bauthor{\bsnm{Kilbourne}, \binits{C.}},
\bauthor{\bsnm{Korpela}, \binits{S.}},
\bauthor{\bsnm{Macculi}, \binits{C.}},
\bauthor{\bsnm{Mendez}, \binits{M.}},
\bauthor{\bsnm{Mitsuda}, \binits{K.}},
\bauthor{\bsnm{Molendi}, \binits{S.}},
\bauthor{\bsnm{Pajot}, \binits{F.}},
\bauthor{\bsnm{Pointecouteau}, \binits{E.}},
\bauthor{\bsnm{Porter}, \binits{F.}},
\bauthor{\bsnm{Pratt}, \binits{G.W.}},
\bauthor{\bsnm{Pr\^ele}, \binits{D.}},
\bauthor{\bsnm{Ravera}, \binits{L.}}:
\batitle{The athena x-ray integral field unit: a consolidated design for the system requirement review of the preliminary definition phase}.
\bjtitle{Exp. Astron.}
\bvolume{56},
\bfpage{213}--\blpage{266}
(\byear{2023})
\doiurl{10.1007/s10686-022-09880-7}
\end{barticle}
\endbibitem

\bibitem[\protect\citeauthoryear{{European Space Agency}}{2024}]{ESA_NewAthena}
\begin{botherref}
\oauthor{\bsnm{{European Space Agency}}}:
{ATHENA Home - NewAthena Mission Overview}.
\url{https://www.cosmos.esa.int/web/athena}.
{Accessed: 2025-07-20}
(2024)
\end{botherref}
\endbibitem

\bibitem[\protect\citeauthoryear{{Kislat} et~al.}{2023}]{kislat_etal_2023}
\begin{barticle}
\bauthor{\bsnm{{Kislat}}, \binits{F.}},
\bauthor{\bsnm{{Becker}}, \binits{D.}},
\bauthor{\bsnm{{Bennett}}, \binits{D.}},
\bauthor{\bsnm{{Dasgupta}}, \binits{A.}},
\bauthor{\bsnm{{Fowler}}, \binits{J.}},
\bauthor{\bsnm{{Fryer}}, \binits{C.L.}},
\bauthor{\bsnm{{Gard}}, \binits{J.}},
\bauthor{\bsnm{{Gau}}, \binits{E.}},
\bauthor{\bsnm{{Gurgew}}, \binits{D.}},
\bauthor{\bsnm{{Harmon}}, \binits{K.}},
\bauthor{\bsnm{{Hayashi}}, \binits{T.}},
\bauthor{\bsnm{{Heatwole}}, \binits{S.}},
\bauthor{\bsnm{{Hossen}}, \binits{M.A.}},
\bauthor{\bsnm{{Krawczynski}}, \binits{H.}},
\bauthor{\bsnm{{Lanzi}}, \binits{R.J.}},
\bauthor{\bsnm{{Legere}}, \binits{J.}},
\bauthor{\bsnm{{Mates}}, \binits{J.A.B.}},
\bauthor{\bsnm{{McConnell}}, \binits{M.}},
\bauthor{\bsnm{{Nagy}}, \binits{J.}},
\bauthor{\bsnm{{Okajima}}, \binits{T.}},
\bauthor{\bsnm{{Sato}}, \binits{T.}},
\bauthor{\bsnm{{Schmidt}}, \binits{D.}},
\bauthor{\bsnm{{Spooner}}, \binits{S.}},
\bauthor{\bsnm{{Swetz}}, \binits{D.}},
\bauthor{\bsnm{{Tamura}}, \binits{K.}},
\bauthor{\bsnm{{Ullom}}, \binits{J.}},
\bauthor{\bsnm{{Weber}}, \binits{J.}},
\bauthor{\bsnm{{Wester}}, \binits{A.}},
\bauthor{\bsnm{{Young}}, \binits{P.}}:
\batitle{{ASCENT: a balloon-borne hard x-ray imaging spectroscopy telescope using transition edge sensor microcalorimeter detectors}}.
\bjtitle{Journal of Astronomical Telescopes, Instruments, and Systems}
\bvolume{9},
\bfpage{014002}
(\byear{2023})
\doiurl{10.1117/1.JATIS.9.1.014002}
{\href{https://arxiv.org/abs/2301.01525}{{arXiv:2301.01525}}}
{[astro-ph.IM]}
\end{barticle}
\endbibitem

\bibitem[\protect\citeauthoryear{{Shirazi} et~al.}{2023}]{shirazi_etal_2023}
\begin{barticle}
\bauthor{\bsnm{{Shirazi}}, \binits{F.}},
\bauthor{\bsnm{{Gau}}, \binits{E.}},
\bauthor{\bsnm{{Hossen}}, \binits{M.A.}},
\bauthor{\bsnm{{Becker}}, \binits{D.}},
\bauthor{\bsnm{{Schmidt}}, \binits{D.}},
\bauthor{\bsnm{{Swetz}}, \binits{D.}},
\bauthor{\bsnm{{Bennett}}, \binits{D.}},
\bauthor{\bsnm{{Braun}}, \binits{D.}},
\bauthor{\bsnm{{Kislat}}, \binits{F.}},
\bauthor{\bsnm{{Gard}}, \binits{J.}},
\bauthor{\bsnm{{Mates}}, \binits{J.}},
\bauthor{\bsnm{{Weber}}, \binits{J.}},
\bauthor{\bsnm{{Rodriguez Cavero}}, \binits{N.}},
\bauthor{\bsnm{{Chun}}, \binits{S.}},
\bauthor{\bsnm{{Lisalda}}, \binits{L.}},
\bauthor{\bsnm{{West}}, \binits{A.}},
\bauthor{\bsnm{{Dev}}, \binits{B.}},
\bauthor{\bsnm{{Ferrer}}, \binits{F.}},
\bauthor{\bsnm{{Bose}}, \binits{R.}},
\bauthor{\bsnm{{Ullom}}, \binits{J.}},
\bauthor{\bsnm{{Krawczynski}}, \binits{H.}}:
\batitle{{511-CAM mission: a pointed 511 keV gamma-ray telescope with a focal plane detector made of stacked transition edge sensor microcalorimeter arrays}}.
\bjtitle{J. Astron. Teles. Instrum. Syst.}
\bvolume{9},
\bfpage{024006}
(\byear{2023})
\doiurl{10.1117/1.JATIS.9.2.024006}
{\href{https://arxiv.org/abs/2206.14652}{{arXiv:2206.14652}}}
{[astro-ph.IM]}
\end{barticle}
\endbibitem

\bibitem[\protect\citeauthoryear{{Iyer} et~al.}{2023}]{iyer_etal_2023}
\begin{barticle}
\bauthor{\bsnm{{Iyer}}, \binits{N.K.}},
\bauthor{\bsnm{{Kiss}}, \binits{M.}},
\bauthor{\bsnm{{Pearce}}, \binits{M.}},
\bauthor{\bsnm{{Stana}}, \binits{T.-A.}},
\bauthor{\bsnm{{Awaki}}, \binits{H.}},
\bauthor{\bsnm{{Bose}}, \binits{R.G.}},
\bauthor{\bsnm{{Dasgupta}}, \binits{A.}},
\bauthor{\bsnm{{De Geronimo}}, \binits{G.}},
\bauthor{\bsnm{{Gau}}, \binits{E.}},
\bauthor{\bsnm{{Hakamata}}, \binits{T.}},
\bauthor{\bsnm{{Ishida}}, \binits{M.}},
\bauthor{\bsnm{{Ishiwata}}, \binits{K.}},
\bauthor{\bsnm{{Kamogawa}}, \binits{W.}},
\bauthor{\bsnm{{Kislat}}, \binits{F.}},
\bauthor{\bsnm{{Kitaguchi}}, \binits{T.}},
\bauthor{\bsnm{{Krawczynski}}, \binits{H.}},
\bauthor{\bsnm{{Lisalda}}, \binits{L.}},
\bauthor{\bsnm{{Maeda}}, \binits{Y.}},
\bauthor{\bsnm{{Matsumoto}}, \binits{H.}},
\bauthor{\bsnm{{Miyamoto}}, \binits{A.}},
\bauthor{\bsnm{{Miyazawa}}, \binits{T.}},
\bauthor{\bsnm{{Mizuno}}, \binits{T.}},
\bauthor{\bsnm{{Rauch}}, \binits{B.F.}},
\bauthor{\bsnm{{Cavero}}, \binits{N.R.}},
\bauthor{\bsnm{{Sakamoto}}, \binits{N.}},
\bauthor{\bsnm{{Sato}}, \binits{J.}},
\bauthor{\bsnm{{Spooner}}, \binits{S.}},
\bauthor{\bsnm{{Takahashi}}, \binits{H.}},
\bauthor{\bsnm{{Takeo}}, \binits{M.}},
\bauthor{\bsnm{{Tamagawa}}, \binits{T.}},
\bauthor{\bsnm{{Uchida}}, \binits{Y.}},
\bauthor{\bsnm{{West}}, \binits{A.T.}},
\bauthor{\bsnm{{Wimalasena}}, \binits{K.}},
\bauthor{\bsnm{{Yoshimoto}}, \binits{M.}}:
\batitle{{The design and performance of the XL-Calibur anticoincidence shield}}.
\bjtitle{Nucl. Instrum. Meth. A}
\bvolume{1048},
\bfpage{167975}
(\byear{2023})
\doiurl{10.1016/j.nima.2022.167975}
{\href{https://arxiv.org/abs/2212.04139}{{arXiv:2212.04139}}}
{[astro-ph.IM]}
\end{barticle}
\endbibitem

\bibitem[\protect\citeauthoryear{{Metzler} et~al.}{2024}]{Metzler:2024dgk}
\begin{bchapter}
\bauthor{\bsnm{{Metzler}}, \binits{Z.}},
\bauthor{\bsnm{{Cannady}}, \binits{N.}},
\bauthor{\bsnm{{Shy}}, \binits{D.}},
\bauthor{\bsnm{{Caputo}}, \binits{R.}},
\bauthor{\bsnm{{Kierans}}, \binits{C.}},
\bauthor{\bsnm{{Woolf}}, \binits{R.}}:
\bctitle{{The anti-coincidence detector subsystem for ComPair}}.
In: \beditor{\bsnm{{den Herder}}, \binits{J.-W.A.}},
\beditor{\bsnm{{Nikzad}}, \binits{S.}},
\beditor{\bsnm{{Nakazawa}}, \binits{K.}} (eds.)
\bbtitle{Space Telescopes and Instrumentation 2024: Ultraviolet to Gamma Ray}.
\bsertitle{Society of Photo-Optical Instrumentation Engineers (SPIE) Conference Series},
vol. \bseriesno{13093},
p. \bfpage{130937}
(\byear{2024}).
\doiurl{10.1117/12.3017653}
\end{bchapter}
\endbibitem

\bibitem[\protect\citeauthoryear{{D'Andrea} et~al.}{2024}]{d-andrea_etal_2024}
\begin{barticle}
\bauthor{\bsnm{{D'Andrea}}, \binits{M.}},
\bauthor{\bsnm{{Macculi}}, \binits{C.}},
\bauthor{\bsnm{{Lotti}}, \binits{S.}},
\bauthor{\bsnm{{Piro}}, \binits{L.}},
\bauthor{\bsnm{{Argan}}, \binits{A.}},
\bauthor{\bsnm{{Minervini}}, \binits{G.}},
\bauthor{\bsnm{{Torrioli}}, \binits{G.}},
\bauthor{\bsnm{{Chiarello}}, \binits{F.}},
\bauthor{\bsnm{{Ferrari Barusso}}, \binits{L.}},
\bauthor{\bsnm{{Celasco}}, \binits{E.}},
\bauthor{\bsnm{{Gallucci}}, \binits{G.}},
\bauthor{\bsnm{{Gatti}}, \binits{F.}},
\bauthor{\bsnm{{Grosso}}, \binits{D.}},
\bauthor{\bsnm{{Rigano}}, \binits{M.}},
\bauthor{\bsnm{{Brienza}}, \binits{D.}},
\bauthor{\bsnm{{Cavazzuti}}, \binits{E.}},
\bauthor{\bsnm{{Volpe}}, \binits{A.}}:
\batitle{{The TES-based Cryogenic AntiCoincidence Detector (CryoAC) of ATHENA X-IFU: A Large Area Silicon Microcalorimeter for Background Particles Detection}}.
\bjtitle{Journal of Low Temperature Physics}
\bvolume{214}(\bissue{3-4}),
\bfpage{164}--\blpage{172}
(\byear{2024})
\doiurl{10.1007/s10909-023-03034-5}
{\href{https://arxiv.org/abs/2401.10827}{{arXiv:2401.10827}}}
{[astro-ph.IM]}
\end{barticle}
\endbibitem

\bibitem[\protect\citeauthoryear{{Murray} et~al.}{2000}]{Murray2000}
\begin{bchapter}
\bauthor{\bsnm{{Murray}}, \binits{S.S.}},
\bauthor{\bsnm{{Chappell}}, \binits{J.H.}},
\bauthor{\bsnm{{Kenter}}, \binits{A.T.}},
\bauthor{\bsnm{{Juda}}, \binits{M.}},
\bauthor{\bsnm{{Kraft}}, \binits{R.P.}},
\bauthor{\bsnm{{Zombeck}}, \binits{M.V.}},
\bauthor{\bsnm{{Meehan}}, \binits{G.R.}},
\bauthor{\bsnm{{Austin}}, \binits{G.K.}},
\bauthor{\bsnm{{Gomes}}, \binits{J.J.}}:
\bctitle{{Event screening for the Chandra X-Ray Observatory High-Resolution Camera (HRC)}}.
In: \beditor{\bsnm{{Flanagan}}, \binits{K.A.}},
\beditor{\bsnm{{Siegmund}}, \binits{O.H.}} (eds.)
\bbtitle{X-Ray and Gamma-Ray Instrumentation for Astronomy XI}.
\bsertitle{Society of Photo-Optical Instrumentation Engineers (SPIE) Conference Series},
vol. \bseriesno{4140},
pp. \bfpage{144}--\blpage{154}
(\byear{2000}).
\doiurl{10.1117/12.409108}
\end{bchapter}
\endbibitem

\bibitem[\protect\citeauthoryear{{Vedrenne} et~al.}{2003}]{SPI-ACS}
\begin{botherref}
\oauthor{\bsnm{{Vedrenne}}, \binits{G.}},
\oauthor{\bsnm{{Roques}}, \binits{J.-P.}},
\oauthor{\bsnm{{Sch{\"o}nfelder}}, \binits{V.}},
\oauthor{\bsnm{{Mandrou}}, \binits{P.}},
\oauthor{\bsnm{{Lichti}}, \binits{G.G.}},
\oauthor{\bsnm{{von Kienlin}}, \binits{A.}},
\oauthor{\bsnm{{Cordier}}, \binits{B.}},
\oauthor{\bsnm{{Schanne}}, \binits{S.}},
\oauthor{\bsnm{{Kn{\"o}dlseder}}, \binits{J.}},
\oauthor{\bsnm{{Skinner}}, \binits{G.}},
\oauthor{\bsnm{{Jean}}, \binits{P.}},
\oauthor{\bsnm{{Sanchez}}, \binits{F.}},
\oauthor{\bsnm{{Caraveo}}, \binits{P.}},
\oauthor{\bsnm{{Teegarden}}, \binits{B.}},
\oauthor{\bsnm{{von Ballmoos}}, \binits{P.}},
\oauthor{\bsnm{{Bouchet}}, \binits{L.}},
\oauthor{\bsnm{{Paul}}, \binits{P.}},
\oauthor{\bsnm{{Matteson}}, \binits{J.}},
\oauthor{\bsnm{{Boggs}}, \binits{S.}},
\oauthor{\bsnm{{Wunderer}}, \binits{C.}},
\oauthor{\bsnm{{Leleux}}, \binits{P.}},
\oauthor{\bsnm{{Weidenspointner}}, \binits{G.}},
\oauthor{\bsnm{{Durouchoux}}, \binits{P.}},
\oauthor{\bsnm{{Diehl}}, \binits{R.}},
\oauthor{\bsnm{{Strong}}, \binits{A.}},
\oauthor{\bsnm{{Cass{\'e}}}, \binits{M.}},
\oauthor{\bsnm{{Clair}}, \binits{M.A.}},
\oauthor{\bsnm{{Andr{\'e}}}, \binits{Y.}}:
{SPI: The spectrometer aboard INTEGRAL}
\textbf{411},
63--70
(2003)
\doiurl{10.1051/0004-6361:20031482}
\end{botherref}
\endbibitem

\bibitem[\protect\citeauthoryear{{Marini Bettolo} et~al.}{2008}]{MariniBettolo2008}
\begin{bchapter}
\bauthor{\bsnm{{Marini Bettolo}}, \binits{C.}},
\bauthor{\bsnm{{Pearce}}, \binits{M.}},
\bauthor{\bsnm{{Kiss}}, \binits{M.}},
\bauthor{\bsnm{{Klamra}}, \binits{W.}},
\bauthor{\bsnm{{Siegl}}, \binits{M.}}:
\bctitle{{The BGO anticoincidence system of the PoGOLite balloon-borne soft gamma-ray polarimeter}}.
In: \bbtitle{International Cosmic Ray Conference}.
\bsertitle{International Cosmic Ray Conference},
vol. \bseriesno{2},
pp. \bfpage{483}--\blpage{486}
(\byear{2008})
\end{bchapter}
\endbibitem

\bibitem[\protect\citeauthoryear{Grefenstette et~al.}{2022}]{Grefenstette2022}
\begin{barticle}
\bauthor{\bsnm{Grefenstette}, \binits{B.W.}},
\bauthor{\bsnm{Madsen}, \binits{K.K.}},
\bauthor{\bsnm{Miyasaka}, \binits{H.}},
\bauthor{\bsnm{Craig}, \binits{W.W.}},
\bauthor{\bsnm{Harrison}, \binits{F.A.}},
\bauthor{\bsnm{Rana}, \binits{V.M.}},
\bauthor{\bsnm{Garcia}, \binits{J.A.}},
\bauthor{\bsnm{Fryer}, \binits{C.L.}},
\bauthor{\bsnm{Boggs}, \binits{S.E.}},
\bauthor{\bsnm{Christensen}, \binits{F.E.}},
\bauthor{\bsnm{Forster}, \binits{K.}},
\bauthor{\bsnm{Hailey}, \binits{C.J.}},
\bauthor{\bsnm{Stern}, \binits{D.}}:
\batitle{{NuSTAR non-x-ray background}}.
\bjtitle{J. Astron. Telesc. Instrum. Syst.}
\bvolume{8}(\bissue{4}),
\bfpage{047001}
(\byear{2022})
\doiurl{10.1117/1.JATIS.8.4.047001}
\end{barticle}
\endbibitem

\bibitem[\protect\citeauthoryear{Zhang et~al.}{2025}]{HERD_ACS}
\begin{barticle}
\bauthor{\bsnm{Zhang}, \binits{Y.}},
\bauthor{\bsnm{Liu}, \binits{Y.}},
\bauthor{\bsnm{Han}, \binits{J.}},
\bauthor{\bsnm{Guo}, \binits{D.}},
\bauthor{\bsnm{Dong}, \binits{Y.}},
\bauthor{\bsnm{Gao}, \binits{M.}},
\bauthor{\bsnm{Fan}, \binits{R.}},
\bauthor{\bsnm{Tan}, \binits{Z.}},
\bauthor{\bsnm{Wang}, \binits{Z.}}:
\batitle{{Radiation characterization of SiPMs for HERD PSD}}.
\bjtitle{Nucl. Instrum. Meth. A}
\bvolume{1070},
\bfpage{170035}
(\byear{2025})
\doiurl{10.1016/j.nima.2024.170035}
\end{barticle}
\endbibitem

\bibitem[\protect\citeauthoryear{{Acerbi} et~al.}{2017}]{FBK}
\begin{barticle}
\bauthor{\bsnm{{Acerbi}}, \binits{F.}},
\bauthor{\bsnm{{Davini}}, \binits{S.}},
\bauthor{\bsnm{{Ferri}}, \binits{A.}},
\bauthor{\bsnm{{Galbiati}}, \binits{C.}},
\bauthor{\bsnm{{Giovanetti}}, \binits{G.}},
\bauthor{\bsnm{{Gola}}, \binits{A.}},
\bauthor{\bsnm{{Korga}}, \binits{G.}},
\bauthor{\bsnm{{Mandarano}}, \binits{A.}},
\bauthor{\bsnm{{Marcante}}, \binits{M.}},
\bauthor{\bsnm{{Paternoster}}, \binits{G.}},
\bauthor{\bsnm{{Piemonte}}, \binits{C.}},
\bauthor{\bsnm{{Razeto}}, \binits{A.}},
\bauthor{\bsnm{{Regazzoni}}, \binits{V.}},
\bauthor{\bsnm{{Sablone}}, \binits{D.}},
\bauthor{\bsnm{{Savarese}}, \binits{C.}},
\bauthor{\bsnm{{Zappal{\'a}}}, \binits{G.}},
\bauthor{\bsnm{{Zorzi}}, \binits{N.}}:
\batitle{{Cryogenic Characterization of FBK HD Near-UV Sensitive SiPMs}}.
\bjtitle{IEEE Transactions on Electron Devices}
\bvolume{64}(\bissue{2}),
\bfpage{521}--\blpage{526}
(\byear{2017})
\doiurl{10.1109/TED.2016.2641586}
{\href{https://arxiv.org/abs/1610.01915}{{arXiv:1610.01915}}}
{[physics.ins-det]}
\end{barticle}
\endbibitem

\bibitem[\protect\citeauthoryear{{Gu} et~al.}{2023}]{2023arXiv231110497G}
\begin{botherref}
\oauthor{\bsnm{{Gu}}, \binits{F.}},
\oauthor{\bsnm{{Liao}}, \binits{J.}},
\oauthor{\bsnm{{Zhou}}, \binits{J.}},
\oauthor{\bsnm{{Ma}}, \binits{M.}},
\oauthor{\bsnm{{Gao}}, \binits{Y.}},
\oauthor{\bsnm{{Peng}}, \binits{Z.}},
\oauthor{\bsnm{{Zheng}}, \binits{J.}},
\oauthor{\bsnm{{An}}, \binits{G.}},
\oauthor{\bsnm{{Zhang}}, \binits{L.}},
\oauthor{\bsnm{{Zhang}}, \binits{L.}},
\oauthor{\bsnm{{Liang}}, \binits{Z.}},
\oauthor{\bsnm{{Zhao}}, \binits{X.}},
\oauthor{\bsnm{{Acerbi}}, \binits{F.}},
\oauthor{\bsnm{{Ficorella}}, \binits{A.}},
\oauthor{\bsnm{{Gola}}, \binits{A.}},
\oauthor{\bsnm{{Parellada Monreal}}, \binits{L.}}:
{Characterization of FBK NUV-HD-Cryo SiPMs near LHe temperature}.
arXiv e-prints,
2311--10497
(2023)
\doiurl{10.48550/arXiv.2311.10497}
{\href{https://arxiv.org/abs/2311.10497}{{arXiv:2311.10497}}}
{[physics.ins-det]}
\end{botherref}
\endbibitem

\bibitem[\protect\citeauthoryear{Gironnet et~al.}{2008}]{BGO_temp}
\begin{barticle}
\bauthor{\bsnm{Gironnet}, \binits{J.}},
\bauthor{\bsnm{Mikhailik}, \binits{V.B.}},
\bauthor{\bsnm{Kraus}, \binits{H.}},
\bauthor{\bsnm{{de Marcillac}}, \binits{P.}},
\bauthor{\bsnm{Coron}, \binits{N.}}:
\batitle{{Scintillation studies of \ce{Bi4Ge3O12} (BGO) down to a temperature of 6K}}.
\bjtitle{Nucl. Instrum. Meth. A}
\bvolume{594}(\bissue{3}),
\bfpage{358}--\blpage{361}
(\byear{2008})
\doiurl{10.1016/j.nima.2008.07.008}
\end{barticle}
\endbibitem

\bibitem[\protect\citeauthoryear{Verdier et~al.}{2011}]{PhysRevB.84.214306}
\begin{barticle}
\bauthor{\bsnm{Verdier}, \binits{M.-A.}},
\bauthor{\bsnm{Di~Stefano}, \binits{P.C.F.}},
\bauthor{\bsnm{Nadeau}, \binits{P.}},
\bauthor{\bsnm{Behan}, \binits{C.}},
\bauthor{\bsnm{Clavel}, \binits{M.}},
\bauthor{\bsnm{Dujardin}, \binits{C.}}:
\batitle{Scintillation properties of bi${}_{4}$ge${}_{3}$o${}_{12}$ down to 3 k under $\ensuremath{\gamma}$ rays}.
\bjtitle{Phys. Rev. B}
\bvolume{84},
\bfpage{214306}
(\byear{2011})
\doiurl{10.1103/PhysRevB.84.214306}
\end{barticle}
\endbibitem

\bibitem[\protect\citeauthoryear{Swiderski et~al.}{2019}]{SWIDERSKI201932}
\begin{barticle}
\bauthor{\bsnm{Swiderski}, \binits{L.}},
\bauthor{\bsnm{Moszyński}, \binits{M.}},
\bauthor{\bsnm{Czarnacki}, \binits{W.}},
\bauthor{\bsnm{Brylew}, \binits{K.}},
\bauthor{\bsnm{Grodzicka-Kobylka}, \binits{M.}},
\bauthor{\bsnm{Mianowska}, \binits{Z.}},
\bauthor{\bsnm{Sworobowicz}, \binits{T.}},
\bauthor{\bsnm{Syntfeld-Każuch}, \binits{A.}},
\bauthor{\bsnm{Szczesniak}, \binits{T.}},
\bauthor{\bsnm{Klamra}, \binits{W.}},
\bauthor{\bsnm{Williams}, \binits{R.T.}},
\bauthor{\bsnm{Gridin}, \binits{S.}},
\bauthor{\bsnm{Lu}, \binits{X.}},
\bauthor{\bsnm{Mayhugh}, \binits{M.R.}},
\bauthor{\bsnm{Gektin}, \binits{A.}},
\bauthor{\bsnm{Vasyukov}, \binits{S.}},
\bauthor{\bsnm{Piemonte}, \binits{C.}},
\bauthor{\bsnm{Acerbi}, \binits{F.}},
\bauthor{\bsnm{Ferri}, \binits{A.}},
\bauthor{\bsnm{Gola}, \binits{A.}},
\bauthor{\bsnm{Zawistowski}, \binits{T.}}:
\batitle{Scintillation response to gamma-rays measured at wide temperature range for tl doped csi with sipm readout}.
\bjtitle{Nuclear Instruments and Methods in Physics Research Section A: Accelerators, Spectrometers, Detectors and Associated Equipment}
\bvolume{916},
\bfpage{32}--\blpage{36}
(\byear{2019})
\doiurl{10.1016/j.nima.2018.10.149}
\end{barticle}
\endbibitem

\bibitem[\protect\citeauthoryear{{Dilillo} et~al.}{2022}]{dilillo_etal_2022}
\begin{barticle}
\bauthor{\bsnm{{Dilillo}}, \binits{G.}},
\bauthor{\bsnm{{Zampa}}, \binits{N.}},
\bauthor{\bsnm{{Campana}}, \binits{R.}},
\bauthor{\bsnm{{Fuschino}}, \binits{F.}},
\bauthor{\bsnm{{Pauletta}}, \binits{G.}},
\bauthor{\bsnm{{Rashevskaya}}, \binits{I.}},
\bauthor{\bsnm{{Ambrosino}}, \binits{F.}},
\bauthor{\bsnm{{Baruzzo}}, \binits{M.}},
\bauthor{\bsnm{{Cauz}}, \binits{D.}},
\bauthor{\bsnm{{Cirrincione}}, \binits{D.}},
\bauthor{\bsnm{{Citossi}}, \binits{M.}},
\bauthor{\bsnm{{Casa}}, \binits{G.D.}},
\bauthor{\bsnm{{Di Ruzza}}, \binits{B.}},
\bauthor{\bsnm{{Evangelista}}, \binits{Y.}},
\bauthor{\bsnm{{Galg{\'o}czi}}, \binits{G.}},
\bauthor{\bsnm{{Labanti}}, \binits{C.}},
\bauthor{\bsnm{{Ripa}}, \binits{J.}},
\bauthor{\bsnm{{Tommasino}}, \binits{F.}},
\bauthor{\bsnm{{Verroi}}, \binits{E.}},
\bauthor{\bsnm{{Fiore}}, \binits{F.}},
\bauthor{\bsnm{{Vacchi}}, \binits{A.}}:
\batitle{{Space applications of GAGG:Ce scintillators: a study of afterglow emission by proton irradiation}}.
\bjtitle{Nucl. Instrum. Meth. B}
\bvolume{513},
\bfpage{33}--\blpage{43}
(\byear{2022})
\doiurl{10.1016/j.nimb.2021.12.006}
{\href{https://arxiv.org/abs/2112.02897}{{arXiv:2112.02897}}}
{[astro-ph.IM]}
\end{barticle}
\endbibitem

\bibitem[\protect\citeauthoryear{{Yoneyama} et~al.}{2018}]{yoneyama_etal_2018}
\begin{barticle}
\bauthor{\bsnm{{Yoneyama}}, \binits{M.}},
\bauthor{\bsnm{{Kataoka}}, \binits{J.}},
\bauthor{\bsnm{{Arimoto}}, \binits{M.}},
\bauthor{\bsnm{{Masuda}}, \binits{T.}},
\bauthor{\bsnm{{Yoshino}}, \binits{M.}},
\bauthor{\bsnm{{Kamada}}, \binits{K.}},
\bauthor{\bsnm{{Yoshikawa}}, \binits{A.}},
\bauthor{\bsnm{{Sato}}, \binits{H.}},
\bauthor{\bsnm{{Usuki}}, \binits{Y.}}:
\batitle{{Evaluation of GAGG:Ce scintillators for future space applications}}.
\bjtitle{J. Instrum.}
\bvolume{13}(\bissue{2}),
\bfpage{02023}
(\byear{2018})
\doiurl{10.1088/1748-0221/13/02/P02023}
\end{barticle}
\endbibitem

\bibitem[\protect\citeauthoryear{Kamada et~al.}{2012}]{Kamada2012}
\begin{barticle}
\bauthor{\bsnm{Kamada}, \binits{K.}},
\bauthor{\bsnm{Yanagida}, \binits{T.}},
\bauthor{\bsnm{Endo}, \binits{T.}},
\bauthor{\bsnm{Tsutumi}, \binits{K.}},
\bauthor{\bsnm{Usuki}, \binits{Y.}},
\bauthor{\bsnm{Nikl}, \binits{M.}},
\bauthor{\bsnm{Fujimoto}, \binits{Y.}},
\bauthor{\bsnm{Fukabori}, \binits{A.}},
\bauthor{\bsnm{Yoshikawa}, \binits{A.}}:
\batitle{2inch diameter single crystal growth and scintillation properties of ce:gd3al2ga3o12}.
\bjtitle{Journal of Crystal Growth}
\bvolume{352}(\bissue{1}),
\bfpage{88}--\blpage{90}
(\byear{2012})
\doiurl{10.1016/j.jcrysgro.2011.11.085} .
\bcomment{The Proceedings of the 18th American Conference on Crystal Growth and Epitaxy}
\end{barticle}
\endbibitem

\bibitem[\protect\citeauthoryear{Park et~al.}{2023}]{Park2023ChemPolish}
\begin{barticle}
\bauthor{\bsnm{Park}, \binits{S.}},
\bauthor{\bsnm{Jeong}, \binits{M.}},
\bauthor{\bsnm{Kim}, \binits{J.}},
\bauthor{\bsnm{Jung}, \binits{J.}},
\bauthor{\bsnm{Lee}, \binits{H.}},
\bauthor{\bsnm{Choi}, \binits{H.}},
\bauthor{\bsnm{Kim}, \binits{J.}},
\bauthor{\bsnm{Lim}, \binits{J.}},
\bauthor{\bsnm{Kim}, \binits{J.}},
\bauthor{\bsnm{Kim}, \binits{J.}},
\bauthor{\bsnm{Lee}, \binits{H.}},
\bauthor{\bsnm{Lee}, \binits{S.}},
\bauthor{\bsnm{Lee}, \binits{J.}},
\bauthor{\bsnm{Kim}, \binits{H.}},
\bauthor{\bsnm{Park}, \binits{J.}}:
\batitle{{Scintillation characteristics of chemically processed Ce:GAGG single crystals}}.
\bjtitle{PLoS ONE}
\bvolume{18}(\bissue{3}),
\bfpage{0281262}
(\byear{2023})
\doiurl{10.1371/journal.pone.0281262}
\end{barticle}
\endbibitem

\bibitem[\protect\citeauthoryear{Park and Jeong}{2023}]{Park2023}
\begin{barticle}
\bauthor{\bsnm{Park}, \binits{S.}},
\bauthor{\bsnm{Jeong}, \binits{M.}}:
\batitle{{Comparison of Characteristics of Gamma-Ray Imager Based on GAGG(Ce) and BGO Scintillators}}.
\bjtitle{J. Radiat. Protection Res.}
\bvolume{48}(\bissue{1}),
\bfpage{25}--\blpage{34}
(\byear{2023})
\doiurl{10.14407/jrpr.2022.00122} .
\bcomment{States BGO light yield is about 5.9 times lower than GAGG(Ce)}
\end{barticle}
\endbibitem

\bibitem[\protect\citeauthoryear{Agostinelli et~al.}{2003}]{Agostinelli2003GEANT4}
\begin{barticle}
\bauthor{\bsnm{Agostinelli}, \binits{S.}},
\bauthor{\bsnm{Allison}, \binits{J.}},
\bauthor{\bsnm{Amako}, \binits{K.}},
\bauthor{\bsnm{Apostolakis}, \binits{J.}},
\bauthor{\bsnm{Araujo}, \binits{H.}}, \betal:
\batitle{{Geant4—a simulation toolkit}}.
\bjtitle{Nucl. Instrum. Meth. A}
\bvolume{506}(\bissue{3}),
\bfpage{250}--\blpage{303}
(\byear{2003})
\doiurl{10.1016/S0168-9002(03)01368-8}
\end{barticle}
\endbibitem

\bibitem[\protect\citeauthoryear{Korzhik et~al.}{2022}]{GAGG_Gd_role}
\begin{botherref}
\oauthor{\bsnm{Korzhik}, \binits{M.}},
\oauthor{\bsnm{Retivov}, \binits{V.}},
\oauthor{\bsnm{Bondarau}, \binits{A.}},
\oauthor{\bsnm{Dosovitskiy}, \binits{G.}},
\oauthor{\bsnm{Dubov}, \binits{V.}},
\oauthor{\bsnm{Kamenskikh}, \binits{I.}},
\oauthor{\bsnm{Karpuk}, \binits{P.}},
\oauthor{\bsnm{Kuznetsova}, \binits{D.}},
\oauthor{\bsnm{Smyslova}, \binits{V.}},
\oauthor{\bsnm{Mechinsky}, \binits{V.}},
\oauthor{\bsnm{Pustovarov}, \binits{V.}},
\oauthor{\bsnm{Tavrunov}, \binits{D.}},
\oauthor{\bsnm{Tishchenko}, \binits{E.}},
\oauthor{\bsnm{Vasil’ev}, \binits{A.}}:
{Role of the Dilution of the Gd Sublattice in Forming the Scintillation Properties of Quaternary (Gd,Lu)3Al2Ga3O12: Ce Ceramics}.
Crystals
\textbf{12}(9)
(2022)
\doiurl{10.3390/cryst12091196}
\end{botherref}
\endbibitem

\bibitem[\protect\citeauthoryear{Drozdowski et~al.}{2014}]{GAGG_traps}
\begin{barticle}
\bauthor{\bsnm{Drozdowski}, \binits{W.}},
\bauthor{\bsnm{Brylew}, \binits{K.}},
\bauthor{\bsnm{Witkowski}, \binits{M.E.}},
\bauthor{\bsnm{Wojtowicz}, \binits{A.J.}},
\bauthor{\bsnm{Solarz}, \binits{P.}},
\bauthor{\bsnm{Kamada}, \binits{K.}},
\bauthor{\bsnm{Yoshikawa}, \binits{A.}}:
\batitle{{Studies of light yield as a function of temperature and low temperature thermoluminescence of $\text{Gd}_3\text{Al}_2\text{Ga}_3\text{O}_{12}\text{:Ce}$ scintillator crystals}}.
\bjtitle{Opt. Mater.}
\bvolume{36}(\bissue{10}),
\bfpage{1665}--\blpage{1669}
(\byear{2014})
\doiurl{10.1016/j.optmat.2013.12.044} .
\bcomment{SI: IWASOM'13}
\end{barticle}
\endbibitem

\bibitem[\protect\citeauthoryear{Zajíc et~al.}{2025}]{GAGG_fast_slow}
\begin{barticle}
\bauthor{\bsnm{Zajíc}, \binits{F.}},
\bauthor{\bsnm{Jarý}, \binits{V.}},
\bauthor{\bsnm{Pospíšil}, \binits{J.}},
\bauthor{\bsnm{Boháček}, \binits{P.}},
\bauthor{\bsnm{Umar}, \binits{Z.}},
\bauthor{\bsnm{Piasecki}, \binits{M.}},
\bauthor{\bsnm{Brik}, \binits{M.G.}},
\bauthor{\bsnm{Kučerková}, \binits{R.}},
\bauthor{\bsnm{Beitlerová}, \binits{A.}},
\bauthor{\bsnm{Nikl}, \binits{M.}}:
\batitle{{A fast GGAG:Ce(Mg) single crystal scintillator: LDFZM growth{,} characterization and electronic band structure calculation}}.
\bjtitle{Mater. Adv.}
\bvolume{6},
\bfpage{777}--\blpage{787}
(\byear{2025})
\doiurl{10.1039/D4MA00976B}
\end{barticle}
\endbibitem

\bibitem[\protect\citeauthoryear{Sibczynski et~al.}{2015}]{GAGG_Al_ratio}
\begin{barticle}
\bauthor{\bsnm{Sibczynski}, \binits{P.}},
\bauthor{\bsnm{Iwanowska-Hanke}, \binits{J.}},
\bauthor{\bsnm{Moszy\'nski}, \binits{M.}},
\bauthor{\bsnm{Swiderski}, \binits{L.}},
\bauthor{\bsnm{Szaw{\l}owski}, \binits{M.}},
\bauthor{\bsnm{Grodzicka}, \binits{M.}},
\bauthor{\bsnm{Szcz\k{e}\'sniak}, \binits{T.}},
\bauthor{\bsnm{Kamada}, \binits{K.}},
\bauthor{\bsnm{Yoshikawa}, \binits{A.}}:
\batitle{{Characterization of GAGG:Ce scintillators with various Al-to-Ga ratio}}.
\bjtitle{Nucl. Instrum. Meth. A}
\bvolume{772},
\bfpage{112}--\blpage{117}
(\byear{2015})
\doiurl{10.1016/j.nima.2014.10.041}
\end{barticle}
\endbibitem

\bibitem[\protect\citeauthoryear{Furuno et~al.}{2021}]{Furuno_2021}
\begin{barticle}
\bauthor{\bsnm{Furuno}, \binits{T.}},
\bauthor{\bsnm{Koshikawa}, \binits{A.}},
\bauthor{\bsnm{Kawabata}, \binits{T.}},
\bauthor{\bsnm{Itoh}, \binits{M.}},
\bauthor{\bsnm{Kurosawa}, \binits{S.}},
\bauthor{\bsnm{Morimoto}, \binits{T.}},
\bauthor{\bsnm{Murata}, \binits{M.}},
\bauthor{\bsnm{Sakanashi}, \binits{K.}},
\bauthor{\bsnm{Tsumura}, \binits{M.}},
\bauthor{\bsnm{Yamaji}, \binits{A.}}:
\batitle{{Response of the GAGG(Ce) scintillator to charged particles compared with the CsI(Tl) scintillator}}.
\bjtitle{J. Instrum.}
\bvolume{16}(\bissue{10}),
\bfpage{10012}
(\byear{2021})
\doiurl{10.1088/1748-0221/16/10/P10012}
\end{barticle}
\endbibitem

\bibitem[\protect\citeauthoryear{Tang et~al.}{2024}]{GAGG_Al_ratio2}
\begin{barticle}
\bauthor{\bsnm{Tang}, \binits{Y.}},
\bauthor{\bsnm{Qiang}, \binits{M.}},
\bauthor{\bsnm{Lou}, \binits{W.}},
\bauthor{\bsnm{Ding}, \binits{Y.}},
\bauthor{\bsnm{Lin}, \binits{H.}},
\bauthor{\bsnm{Hong}, \binits{R.}},
\bauthor{\bsnm{Zhang}, \binits{D.}}:
\batitle{Ga/al ratio induced afterglow behavior of ce3+:gagg scintillation ceramics}.
\bjtitle{Ceramics International}
\bvolume{50}(\bissue{19, Part B}),
\bfpage{36286}--\blpage{36294}
(\byear{2024})
\doiurl{10.1016/j.ceramint.2024.07.012}
\end{barticle}
\endbibitem

\bibitem[\protect\citeauthoryear{Melecio et~al.}{2024}]{GRAPE}
\begin{bchapter}
\bauthor{\bsnm{Melecio}, \binits{K.O.}},
\bauthor{\bsnm{McConnell}, \binits{M.}},
\bauthor{\bsnm{Bundock}, \binits{J.}},
\bauthor{\bsnm{Ertley}, \binits{C.}},
\bauthor{\bsnm{Kislat}, \binits{F.}},
\bauthor{\bsnm{Kole}, \binits{M.}},
\bauthor{\bsnm{Legere}, \binits{J.}},
\bauthor{\bsnm{Mello}, \binits{E.}},
\bauthor{\bsnm{Mello}, \binits{K.}},
\bauthor{\bsnm{Puopolo}, \binits{D.}},
\bauthor{\bsnm{Zaid}, \binits{J.}}:
\bctitle{{Gamma-ray observations at stratospheric altitudes with the Gamma-Ray Polarimeter Experiment (GRAPE)}}.
In: \bbtitle{Space Telescopes and Instrumentation 2024: Ultraviolet to Gamma Ray},
vol. \bseriesno{13093}
(\byear{2024}).
\doiurl{10.1117/12.3020647} .
\burl{https://doi.org/10.1117/12.3020647}
\end{bchapter}
\endbibitem

\bibitem[\protect\citeauthoryear{{Kole} et~al.}{2025}]{Kole2025UniversalSiPM}
\begin{barticle}
\bauthor{\bsnm{{Kole}}, \binits{M.}},
\bauthor{\bsnm{{De Angelis}}, \binits{N.}},
\bauthor{\bsnm{{Produit}}, \binits{N.}},
\bauthor{\bsnm{{Cadoux}}, \binits{F.}},
\bauthor{\bsnm{{Favre}}, \binits{Y.}},
\bauthor{\bsnm{{Greiner}}, \binits{J.}},
\bauthor{\bsnm{{Hulsman}}, \binits{J.}},
\bauthor{\bsnm{{Kusyk}}, \binits{S.}},
\bauthor{\bsnm{{Li}}, \binits{H.}},
\bauthor{\bsnm{{Rybka}}, \binits{D.}},
\bauthor{\bsnm{{Stauffer}}, \binits{J.}},
\bauthor{\bsnm{{Stil}}, \binits{A.}},
\bauthor{\bsnm{{Sun}}, \binits{J.}},
\bauthor{\bsnm{{Swakon}}, \binits{J.}},
\bauthor{\bsnm{{Wrobel}}, \binits{D.}},
\bauthor{\bsnm{{Wu}}, \binits{X.}}:
\batitle{{Design and performance of a universal SiPM readout system for X- and gamma-ray missions}}.
\bjtitle{Nuclear Instruments and Methods in Physics Research A}
\bvolume{1080},
\bfpage{170782}
(\byear{2025})
\doiurl{10.1016/j.nima.2025.170782}
{\href{https://arxiv.org/abs/2501.07758}{{arXiv:2501.07758}}}
{[astro-ph.IM]}
\end{barticle}
\endbibitem

\bibitem[\protect\citeauthoryear{Nakajima et~al.}{2019}]{GAGG_decay}
\begin{barticle}
\bauthor{\bsnm{Nakajima}, \binits{K.}},
\bauthor{\bsnm{Tamagawa}, \binits{Y.}},
\bauthor{\bsnm{Ogawa}, \binits{I.}},
\bauthor{\bsnm{Tomita}, \binits{S.}},
\bauthor{\bsnm{Masuda}, \binits{A.}},
\bauthor{\bsnm{Kobayashi}, \binits{M.}}:
\batitle{{Temperature dependence of scintillation properties and pulse shape discrimination between $\gamma$- and $\alpha$-rays in $\text{Gd}_3\text{Al}_2\text{Ga}_3\text{O}_{12}\text{:Ce}$ scintillator}}.
\bjtitle{Nucl. Instrum. Meth. A}
\bvolume{916},
\bfpage{51}--\blpage{55}
(\byear{2019})
\doiurl{10.1016/j.nima.2018.11.012}
\end{barticle}
\endbibitem

\bibitem[\protect\citeauthoryear{{Hamamatsu Photonics K.K.}}{2022}]{R1924A}
\begin{botherref}
\oauthor{\bsnm{{Hamamatsu Photonics K.K.}}}:
R1924A/R1924P Photomultiplier Tube Specifications.
(2022).
Accessed July 29, 2025.
\url{https://www.hamamatsu.com/content/dam/hamamatsu-photonics/sites/documents/99_SALES_LIBRARY/etd/R1924A_P_TPMH1280E.pdf}
\end{botherref}
\endbibitem

\bibitem[\protect\citeauthoryear{Technology}{2025}]{Eljen}
\begin{botherref}
\oauthor{\bsnm{Technology}, \binits{E.}}:
EJ-200, EJ-204, EJ-208, EJ-212 - Plastic Scintillators.
\url{https://eljentechnology.com/products/plastic-scintillators/ej-200-ej-204-ej-208-ej-212}.
Accessed: 2025-07-31
(2025)
\end{botherref}
\endbibitem

\bibitem[\protect\citeauthoryear{{Ryan} et~al.}{2010}]{ryan_nspect_2010}
\begin{bchapter}
\bauthor{\bsnm{{Ryan}}, \binits{J.M.}},
\bauthor{\bsnm{{Bancroft}}, \binits{C.}},
\bauthor{\bsnm{{Bloser}}, \binits{P.}},
\bauthor{\bsnm{{Bravar}}, \binits{U.}},
\bauthor{\bsnm{{Fourguette}}, \binits{D.}},
\bauthor{\bsnm{{Frost}}, \binits{C.}},
\bauthor{\bsnm{{Larocque}}, \binits{L.}},
\bauthor{\bsnm{{McConnell}}, \binits{M.L.}},
\bauthor{\bsnm{{Legere}}, \binits{J.}},
\bauthor{\bsnm{{Pavlich}}, \binits{J.}},
\bauthor{\bsnm{{Ritter}}, \binits{G.}},
\bauthor{\bsnm{{Wassick}}, \binits{G.}},
\bauthor{\bsnm{{Wood}}, \binits{J.}},
\bauthor{\bsnm{{Woolf}}, \binits{R.}}:
\bctitle{{A portable neutron spectroscope (NSPECT) for detection, imaging and identification of nuclear material}}.
In: \beditor{\bsnm{{Doty}}, \binits{F.P.}},
\beditor{\bsnm{{Barber}}, \binits{H.B.}},
\beditor{\bsnm{{Roehrig}}, \binits{H.}},
\beditor{\bsnm{{Schirato}}, \binits{R.C.}} (eds.)
\bbtitle{Penetrating Radiation Systems and Applications XI}.
\bsertitle{Society of Photo-Optical Instrumentation Engineers (SPIE) Conference Series},
vol. \bseriesno{7806},
p. \bfpage{780607}
(\byear{2010}).
\doiurl{10.1117/12.860652}
\end{bchapter}
\endbibitem

\bibitem[\protect\citeauthoryear{Tokár et~al.}{1999}]{Tokar1999}
\begin{botherref}
\oauthor{\bsnm{Tokár}, \binits{S.}},
\oauthor{\bsnm{Sýkora}, \binits{I.}},
\oauthor{\bsnm{Pikna}, \binits{M.}}:
{Single Photoelectron Spectra Analysis for the Metal Dynode Photomultiplier}.
Technical Report ATL-TILECAL-99-005,
Department of Nuclear Physics, Comenius University, and Joint Institute for Nuclear Research,
Bratislava, Slovakia and Dubna, Russia
(February 1999).
\url{http://cds.cern.ch/record/683800/files/tilecal-99-005.pdf}
\end{botherref}
\endbibitem

\bibitem[\protect\citeauthoryear{{Yong Hee Photonics}}{2025}]{yphotonics154}
\begin{botherref}
\oauthor{\bsnm{{Yong Hee Photonics}}}:
{Product page: Yong Hee Photonics}.
\url{https://www.yphotonics.com/page5?product_id=154}.
Accessed: 2025-07-14
(2025)
\end{botherref}
\endbibitem

\bibitem[\protect\citeauthoryear{{Quilliam} et~al.}{2013}]{GAG}
\begin{barticle}
\bauthor{\bsnm{{Quilliam}}, \binits{J.A.}},
\bauthor{\bsnm{{Meng}}, \binits{S.}},
\bauthor{\bsnm{{Craig}}, \binits{H.A.}},
\bauthor{\bsnm{{Corruccini}}, \binits{L.R.}},
\bauthor{\bsnm{{Balakrishnan}}, \binits{G.}},
\bauthor{\bsnm{{Petrenko}}, \binits{O.A.}},
\bauthor{\bsnm{{Gomez}}, \binits{A.}},
\bauthor{\bsnm{{Kycia}}, \binits{S.W.}},
\bauthor{\bsnm{{Gingras}}, \binits{M.J.P.}},
\bauthor{\bsnm{{Kycia}}, \binits{J.B.}}:
\batitle{{Juxtaposition of spin freezing and long range order in a series of geometrically frustrated antiferromagnetic gadolinium garnets}}.
\bjtitle{Physical Review B}
\bvolume{87}(\bissue{17}),
\bfpage{174421}
(\byear{2013})
\doiurl{10.1103/PhysRevB.87.174421}
{\href{https://arxiv.org/abs/1006.2969}{{arXiv:1006.2969}}}
{[cond-mat.dis-nn]}
\end{barticle}
\endbibitem

\end{thebibliography}

\end{document}